\documentclass[aps, prb, twocolumn]{revtex4-1}

\usepackage{amsmath}
\usepackage{bm}
\usepackage{graphicx}

\def\ie{{\it i.e.}, }
\def\eg{{\it e.g.}, }

\def\vec#1{{\bm{#1}}}
\def\H{{\mathcal{H}}}
\def\L{{\mathcal{L}}}
\def\A{{\mathcal{A}}}
\def\F{{\mathcal{F}}}
\def\op#1{{\widehat{\bm{#1}}}}
\def\cc{{\rm c.c.}}

\begin{document}

\title{Vector Hamiltonian Formalism for Nonlinear Magnetization Dynamics}
\author{Vasyl Tyberkevych$^\dag$, Andrei Slavin$^\dag$, Petro Artemchuk$^\dag$, Graham Rowlands$^\ddag$}
\email{tyberkev@oakland.edu}
\affiliation{$^\dag$ Department of Physics, Oakland University, Rochester, MI 48309, USA}
\affiliation{$^\ddag$ Raytheon BBN Technologies, Cambridge, MA 02138, USA}
\date\today

\begin{abstract}
Vector Hamiltonian formalism (VHF) for the description of a weakly nonlinear magnetization dynamics has been developed. Transformation from the traditional Landau-Lifshitz equation, describing dynamics of a magnetization vector $\vec{m}(\vec{r}, t)$  on a sphere, to a vector Hamiltonian equation, describing dynamics of a \emph{spin excitation vector} $\vec{s}(\vec{r}, t)$ on a plane, is done using the azimuthal Lambert transformation that preserves both the phase-space area and vector structure of dynamical equations, and guarantees that the plane containing vector $\vec{s}(\vec{r}, t)$ is at each value of the coordinate $\vec{r}$ perpendicular to the a stationary vector $\vec{m}_0(\vec{r})$ describing the magnetization ground state of the system. By expanding vector $\vec{s}(\vec{r}, t)$ in a complete set of linear magnetic vector eigemodes $\vec{s}_\nu(\vec{r})$ of the studied system, and using a weakly nonlinear approximation $|\vec{s}(\vec{r}, t)| \ll 1$, it is possible to express the Hamiltonian function of the system in the form of integrals over the vector eigenmode profiles $\vec{s}_\nu(\vec{r})$, and calculate all the coefficients of this Hamiltonian. The developed approach allows one to describe weakly nonlinear dynamics in micro- and nano-scale magnetic systems with complicated geometries and spatially non-uniform ground states by numerically calculating linear spectrum and eigenmode profiles, and semi-analytically evaluating amplitudes of multi-mode nonlinear interactions. Examples of applications of the developed formalism to the magnetic systems having spatially nonuniform ground state of magnetization are presented.
\end{abstract}

\maketitle

\section{Introduction}
\label{s:Intro}

Weakly nonlinear dynamics of waves having different physical nature is strikingly similar. Nonlinear resonance, parametric instabilities, self-interaction and self-focusing leading to the formation of one- and two-dimensional solitons, generation of higher harmonics -- all these effects are common features for the dynamics of weakly nonlinear optical waves, waves in plasma, waves on a liquid surface, spin waves in magnetically ordered materials, etc. Therefore, it is only natural that there were many attempts to develop a generalized theoretical description of weakly nonlinear dynamics of waves, and this description was based on the classical Hamiltonian formalism \cite{Zakharov1974, Zakharov1975, Lvov1994, Zakharov1997, Gelash2017}. The idea of this approach is to find in each particular case a pair of canonically conjugated variables, in terms of which the natural equations of motion describing a particular wave system are transformed into a pair of standard Hamiltonian equations \cite{Landau1966}, while the energy of the system becomes a Hamiltonian function. Furthermore, a canonical transformation to complex canonical variables (see \cite{Landau1966} and section 1.1.2 in \cite{Lvov1994} for details) allows one to replace two real Hamiltonian equations by one complex equation, and leaves a considerable freedom in the actual choice of these complex variables, that now have the same dimension. The Hamiltonian approach was quite productive, and allowed to consider the weakly nonlinear wave process from a general point of view, independently of the physical nature of a particular wave system. 

In this work we are mainly interested in the nonlinear dynamics in the system of spin waves in magnetically ordered materials, and will concentrate on this particular case. The dynamics of the normalized magnetization vector in this case is described by the Landau-Lifshitz equation (LLE) Eq.~(\ref{LLE}) which, naturally, conserves the length of the magnetization vector $|\vec{m}| = 1$ in recognition of a very strong uniform internal exchange magnetic field existing inside a ferromagnetic material. The Hamiltonian approach to this important system was, first, introduced by Schloemann \cite{Schloemann1960} with the help of the Holstein-Primakoff transformation \cite{Holstein1940}, and was further developed in Refs.~\cite{Zakharov1970, Zakharov1975, Lvov1994}. An alternative transformation bringing the vectorial LLE to the complex Hamiltonian form was introduced in Ref.~\cite{Baryakhtar1975}. The Hamiltonian approach developed for spin waves in an unbounded ferromagnet in Refs.~\cite{Zakharov1970, Zakharov1975, Lvov1994} was quite successful. It made possible the calculation of explicit expressions for the spin wave spectrum and for the three-wave and four-wave nonlinear coefficients of spin wave interactions in unbounded ferromagnetic and antiferromagnetic dielectrics. It also provided a quantitative theory describing the parametric excitation of spin waves and nonlinear stage or a weak spin wave turbulence.  

Later, this Hamiltonian formalism was applied for the case of spin waves propagating in ferromagnetic films of a finite thickness \cite{Slavin1987, Slavin1994, Slavin2008, Krivosik2010} where magnetic eigen-excitations are non-uniform spin wave waves that have a discrete spectrum and are described by plane waves in the film plane, but have well-defined distributions of magnetization along the film thickness determined by the boundary conditions for the magnetization at the film surfaces. Very recently this formalism has been extended to include anti-symmetric interactions, such as Dzyaloshinskii-Moriya interaction \cite{Verba2019}.

It should be noted, that the technical calculations of the interaction coefficients entering the spin wave Hamiltonian for magnetic films \cite{Slavin1994, Slavin2008, Krivosik2010, Verba2019} performed in the framework of the classical Hamiltonian formalism for spin waves \cite{Zakharov1975, Lvov1994} are quite cumbersome and technically rather complicated, because in a vectorial LLE Eq.~(\ref{LLE}) it is necessary to  make a transformation to complex scalar canonical variables expressed in terms of Cartesian components of the magnetization vector (see e.g. section 3.4.2 in \cite{Lvov1994} or Eqs.~(3-10)--(3-13) in \cite{Slavin2008}). The situation here is similar to the situation in electrodynamics where, when switching to Cartesian projections of vectors, you get instead of four vector Maxwell equations twelve scalar equations for the projections of electromagnetic field vectors. Also, the standard scalar canonical transformations bringing LLE to a complex Hamiltonian form are applicable only in the case of spatially uniform ground state of a static magnetization. Modification of the standard approach to a spatially-nonuniform magnetization ground state brings in even more technical difficulties and was used only in a few special cases \cite{Abyzov1979}. 

At the same time, the recent progress in nano-magnetism created a necessity to study nonlinear spin wave processes in micro- and nano-sized magnetic samples that can have strongly non-uniform magnetic ground states in the form of magnetic vortices \cite{Guslienko2011, Schultheiss2019} or skyrmions \cite{Garst2017}, states containing well-defined domain walls \cite{Sluka2019}, or non-uniform ground states simply related to a finite lateral size of a magnetic sample \cite{Bayer2005}.
 
Thus, it is highly desirable to develop a modified variant of the Hamiltonian formalism for spin waves, that is more compact by being able to deal directly with the magnetization vector without the necessity to use complex canonical variable composed of the magnetization projections, and, also, capable to deal with the cases when the magnetic ground state of a considered object is spatially non-uniform. The necessity of such an advanced Hamiltonian approach is also supported by the progress in the research in macroscopic quantum phenomena involving magnons \cite{Demokritov2006}, which now has shifted in the direction of investigation of the nonlinear properties \cite{Dzyapko2017, Borisenko2020} and secondary magnetic excitations in the dense magnon gases \cite{Tiberkevich2019} and Bose-Einstein condensates of magnons \cite{Bozhko2019}. Further development of the nonlinear theory of spin wave generation \cite{Slavin2009}, propagation and synchronization \cite{Awad2017} in magnetic nanostructures also will strongly benefit from the introduction of a novel vectorial Hamiltonian formalism for spin waves.

Our current work represents an attempt to develop a vectorial Hamiltonian formalism for spin waves. Since the magnetization dynamics governed by the LLE Eq.~(\ref{LLE}) is dynamics on a sphere of a unit radius, our first goal would be to map this dynamics vectorially (see Eq.~(\ref{mapping})) on a plane tangential to this sphere, and containing  two-dimensional vector of a magnetic excitation $\vec{s}$ which is everywhere orthogonal to the  coordinate-dependent vector $\vec{m}_0$ describing the spatially non-uniform magnetic ground state of the system. This approach allows us to deal only with vector quantities and, eventually, obtain the expressions of all the spin wave interaction coefficients in the relatively simple and compact vectorial form.

\section{Vector Hamiltonian Formalism}
\label{s:VHF}

In this section we consider the core part of the vector Hamiltonian formalism (VHF), namely, weakly nonlinear autonomous conservative dynamics of magnetic excitations. For simplicity, we shall assume that the considered magnetic body has a finite volume $V_s$, which means that the spectrum of eigen-excitations is discrete and spin wave eigen-modes have finite support and finite norms. This assumption is not critical for the developed formalism and, using standard methods of solid-state theory, one can easily adapt it for description of magnetic excitations in infinite systems (\eg plane spin waves in bulk samples or spin wave modes in thin magnetic films). Modifications of the VHF to the case of dissipative (\eg Gilbert damping or spin transfer torque) and/or non-autonomous (\eg excitation of spin waves by a microwave magnetic field) interactions will be considered in Sec.~\ref{s:Perturbation} in the framework of the general perturbation theory.

\subsection{Landau-Lifshits Equation}
\label{ss:LLE}

The starting point in theoretical analysis of any magnetization dynamics problem is the Landau-Lifshits equation (LLE), which can be written as
\begin{equation}\label{LLE}
	\frac{\partial\vec{m}}{\partial t} = \gamma \left( \vec{B}_{\rm eff} \times \vec{m} \right)
\,,\end{equation}
where $\vec{m} \equiv \vec{m}(t, \vec{r})$ is the unit vector along the magnetization direction, $\gamma$ is the modulus of the gyromagnetic ratio, and $\vec{B}_{\rm eff}$ is the effective magnetic field connected with the energy (Hamiltonian) $\H$ of the magnetic system by
\begin{equation}\label{Beff}
	\vec{B}_{\rm eff} = -\frac{1}{M_s}\frac{\delta\H}{\delta\vec{m}}
\,.\end{equation}
Here $M_s$ is the saturation magnetization and $\delta/\delta\vec{m}$ denotes variational derivative with respect to the field $\vec{m}$. Note, that, in a general case, we allow both $\gamma$ and $M_s$ to depend on position $\vec{r}$, \ie the developed formalism can be used for description of spin dynamics in spatially nonuniform magnetic samples or/and magnetic systems composed of several different magnetic materials.

The LLE Eq.~(\ref{LLE}) can be derived using the least action principle from the Lagrangian function
\begin{equation}\label{LLE-Lagrangian}
	\L = \int L_s \left(\frac{d\A}{dt}\right) d\vec{r} - \H
\,,\end{equation}
where
\begin{equation}\label{def-Ls}
	L_s \equiv M_s/\gamma
\end{equation}
is the density of the spin angular momentum (spin density) associated with the magnetization $M_s$ of the magnetic medium and $d\A$ is the area element encircled by the moving vector $\vec{m}$ on the unit sphere:
\begin{equation}\label{def-dA}
	d\A \equiv \frac{\vec{n} \times \vec{m}}{1 + \vec{n} \cdot \vec{m}} \cdot d\vec{m}
\,.\end{equation}
Here $\vec{n}$ is an arbitrary unit vector, possibly position-dependent (but independent of time). Different choices of $\vec{n}$ lead to Lagrangian functions Eq.~(\ref{LLE-Lagrangian}) that differ by a complete time derivative and, therefore, induce the same equation of motion Eq.~(\ref{LLE}).

The first term in the Lagrangian Eq.~(\ref{LLE-Lagrangian}) is the rate of change of phase-space area due to the motion of the magnetization vector $\vec{m}$ and is analogous to the term $p\,\dot{q}$ in a standard phase-space Lagrangian dynamics. The complicated form of the area element Eq.~(\ref{def-dA}) is due to the fact that the phase space of a magnetization vector is a \emph{sphere} rather than a \emph{plane}, as it is for standard Lagrangian and Hamiltonian systems. One can significantly simplify description of dynamics of a magnetic system by projecting spherical phase space of $\vec{m}$ into a plane, which is the main idea of the current work and previous approaches based on classical complex Hamiltonian formalism for magnetization dynamics. Our approach differs from the predecessors by the choice of the projection function (see Sec.~\ref{ss:SEV} below), which, we believe, is much better suited for modern problems in magnetization dynamics.


We shall write the magnetic energy $\H$ of the system in the form
\begin{equation}\label{energy}
	\H = \int_{V_s}\left[ -M_s \, \vec{B}_{\rm ext} \cdot \vec{m} + \frac12 \, \vec{m} \cdot \op{H} \cdot \vec{m} \right] \, d\vec{r}
\,.\end{equation}
Here $\vec{B}_{\rm ext} \equiv \vec{B}_{\rm ext}(\vec{r})$ is the external magnetic field and $\op{H}$ is a certain Hermitian operator describing self-interactions in the system. Most of the common magnetic self-interactions can be written in such form. Thus, the inhomogeneous exchange is described by the operator
\begin{subequations}\label{opH-cases}
\begin{equation}\label{opH-ex}
	\op{H}_{\rm ex} = -\mu_0 M_s^2 \lambda_{\rm ex}^2 \nabla^2
\,,\end{equation}
where $\mu_0$ is the vacuum permeability and $\lambda_{\rm ex} = \sqrt{A/(\mu_0M_s^2)}$ is the exchange length ($A$ is the exchange stiffness). The operator of the dipolar interaction can be written symbolically as
\begin{equation}\label{opH-dip}
	\op{H}_{\rm dip} = \mu_0 M_s^2 \vec{\nabla} \nabla^{-2} \vec{\nabla}
\,.\end{equation}
In the case of easy-axis uniaxial anisotropy with anisotropy axis $\vec{n}_{\rm an}$ and effective field $B_{\rm an} = 2K_u/M_s$ ($K_u$ is the energy density of uniaxial anisotropy) the interaction operator reads
\begin{equation}\label{opH-an}
	\op{H}_{\rm an} = -M_s B_{\rm an} \vec{n}_{\rm an} \otimes \vec{n}_{\rm an}
\,,\end{equation}
where $\otimes$ denotes direct vector product. An easy-plane anisotropy is described by the same expression Eq.~(\ref{opH-an}) with negative field $B_{\rm an} < 0$. Finally, the Dzyaloshinskii-Moriya interaction (DMI), in the most general case, can be described by the tensor operator
\begin{equation}\label{opH-DMI}
	\op{H}_{\rm DMI} = \op{\Gamma}_{\rm DMI} \cdot \vec{\nabla}
\,,\end{equation}
\end{subequations}
where $\op{\Gamma}_{\rm DMI}$ is a certain third-rank tensor. Strictly speaking, Eqs.~(\ref{opH-cases}) are valid only in the usual case of magnetically-uniform medium (for example, one can easily see that the dipolar operator Eq.~(\ref{opH-dip}) is not Hermitian if $M_s$ depends on $\vec{r}$), but their correction for non-uniform case does not represent any difficulties.

The only relatively common magnetic interaction that cannot be described by the bi-linear magnetic energy operator $\op{H}$ is cubic crystallographic anisotropy. Description of cubic anisotropy would require modification of Eq.~(\ref{energy}) to include additional term proportional to $\vec{m}^4$. The proposed formalism can be generalized to such cases without any principal changes, but it would lead to more complicated expressions for all linear and nonlinear coefficients, and we will not consider such cases here.

For the choice of the energy functional Eq.~(\ref{energy}), the effective magnetic field Eq.~(\ref{Beff}) takes the simple form
\begin{subequations}
\begin{equation}\label{Beff-1}
	\vec{B}_{\rm eff}(\vec{m}) = \vec{B}_{\rm ext} - M_s^{-1} \op{H} \cdot \vec{m}
\,,\end{equation}
which is linear in $\vec{m}$ and clearly demonstrates that nonlinearity of the magnetization dynamics in the system Eq.~(\ref{energy}) is connected solely with the curvature of the phase space.
	
One can also rewrite Eq.~(\ref{Beff-1}) as
\begin{equation}\label{Beff-2}
	\op{H} \cdot \vec{m} = -M_s\left[\vec{B}_{\rm eff}(\vec{m}) - \vec{B}_{\rm ext}\right] = -M_s \vec{B}_{\rm eff}^{(0)}(\vec{m})
\,,\end{equation}
\end{subequations}
where $\vec{B}_{\rm eff}^{(0)}(\vec{m})$ is the self-interaction (\ie without external field $\vec{B}_{\rm ext}$) effective magnetic field created by the magnetization distribution $\vec{m}$. This equation shows, that the operator $\op{H}$ can be simply expressed through the effective magnetic field. Note, that all numerical LLE solvers provides means for calculation of the effective field $\vec{B}_{\rm eff}(\vec{m})$ and, respectively, numerical calculation of the action of the operator $\op{H}$ on any magnetization field $\vec{m}$ does not represent any difficulty and does not require any complicated coding.

To proceed with the problem of weakly nonlinear magnetization dynamics, one also has to specify stationary (or ground) magnetization state $\vec{m}_0(\vec{r})$, around which the dynamics occurs. The ground state $\vec{m}_0(\vec{r})$ is a stationary solution of Eq.~(\ref{LLE}) and can be found from the equation
\begin{equation}\label{ground}
	\vec{B}_{\rm eff}(\vec{m}_0) = \vec{B}_{\rm ext} - M_s^{-1} \op{H} \cdot \vec{m}_0 = B_0 \vec{m}_0
\,,\end{equation}
where $B_0 \equiv B_0(\vec{r})$ is the scalar internal magnetic field. In general, a magnetic system can have several ground states for the same set of parameters. In the following, we shall assume that the ground state of interest is known (both $\vec{m}_0$ and $B_0$). We would like to emphasize, that we do not assume that the ground state is spatially-uniform, and the developed formalism can be used for description of spin excitations of highly inhomogeneous magnetic states such as \eg a domain wall, a magnetic vortex, or a magnetic skyrmion.

Note, that the magnetic energy Eq.~(\ref{energy}) can be written as
\begin{eqnarray}\label{energy-1}
	\H(\vec{m}) &&= \H(\vec{m}_0) \\\nonumber&&
		+\frac12 \int_{V_s} (\vec{m} - \vec{m}_0) \cdot (\op{H} + M_s B_0 \op{I}) \cdot (\vec{m} - \vec{m}_0) \, d\vec{r}
\,,\end{eqnarray}
where $\H(\vec{m}_0)$ is the ground state energy, which does not influence the magnetization dynamics and will be ignored in the following, and $\op{I}$ is the identity operator. Eq.~(\ref{energy-1}) shows that the first-order (in magnetization deviation $(\vec{m} - \vec{m}_0)$) contribution to the magnetic energy vanishes near the ground state $\vec{m}_0$. It also shows that the state $\vec{m}_0$ is stable (corresponds to an energy minimum) if the operator $(\op{H} + M_s B_0 \op{I})$ is positive-definite for allowed small deviations $(\vec{m} - \vec{m}_0)$. Below we shall assume that this condition holds unless otherwise stated.

\subsection{Spin Excitation Vector}
\label{ss:SEV}

To simplify the description of the magnetization dynamics, we shall project the spherical phase space of the magnetization vector $\vec{m}$ into a plane. Namely, at every point $\vec{r}$ we project the unit sphere $\vec{m}(t, \vec{r})$ into the plane $\vec{s}(t, \vec{r})$ orthogonal to the ground state $\vec{m}_0(\vec{r})$ at this point ($\vec{m}_0 \cdot \vec{s} \equiv 0$) using the transformation
\begin{equation}\label{mapping}
	\vec{m} = \left(1 - \frac{s^2}{2}\right) \vec{m}_0 + \sqrt{1 - \frac{s^2}{4}} \, \vec{s}
\,.\end{equation}
Here $s = |\vec{s}|$. The vector $\vec{s}(t, \vec{r})$ will be called \emph{spin excitation vector} (SEV) below. As one can easily show, Eq.~(\ref{mapping}) provides conservation of the length of vector $\vec{m}$ ($\vec{m} \cdot \vec{m} = 1$) for any choice of $\vec{s}$ orthogonal to $\vec{m}_0$.

\begin{figure}
	\includegraphics[width=0.45\textwidth]{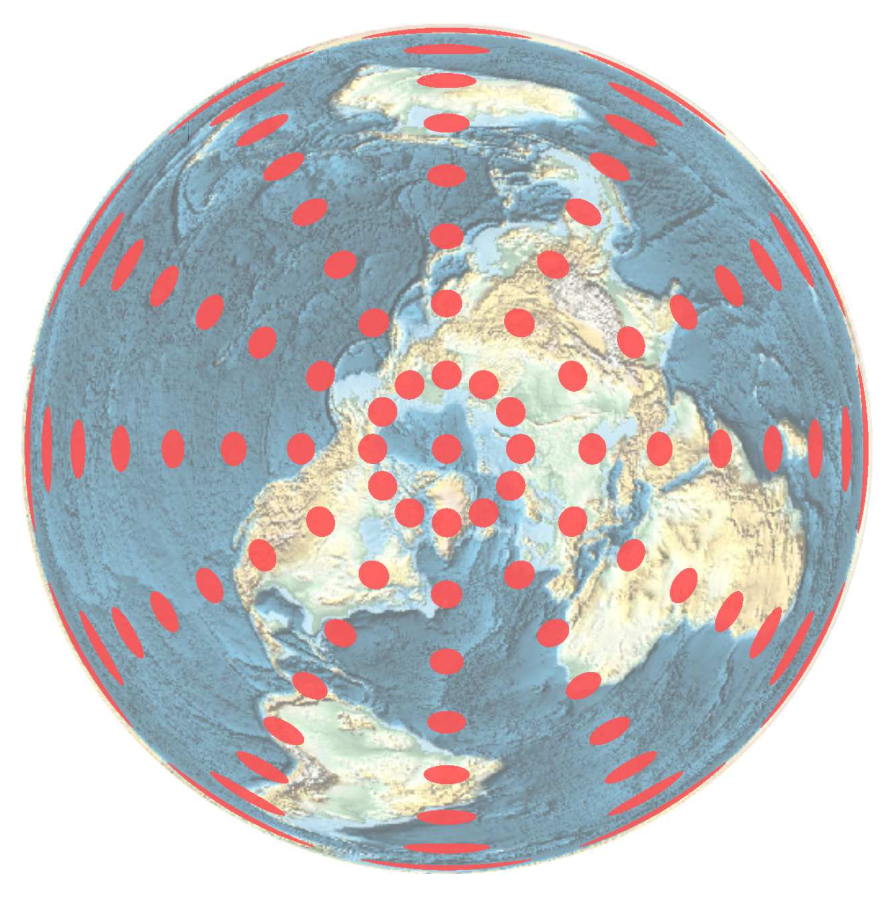}
	\caption{Lambert azimuthal projection Eq.~(\ref{mapping}) of the Earth surface with the ``equilibrium direction'' $\vec{m}_0$ at the North pole. Red ellipses indicate projections of small-size equal-area disks. Note, that noticeable distortions of the disk shapes start only in the southern hemisphere.}
	\label{fig:Lambert}
\end{figure}

The mapping Eq.~(\ref{mapping}) is known in cartography as the Lambert azimuthal equal-area projection (see Fig.~\ref{fig:Lambert}) and has two important properties. First, it is an \emph{equal-area} transformation, \ie it maps an arbitrary region on the sphere $\vec{m}$ into a region in the plane $\vec{s}$ of a different shape, but the same area. The equal-area property ensures that the Hamiltonian nature of magnetization dynamics will be preserved after the transformation to the SEV $\vec{s}$. Second, Eq.~(\ref{mapping}) is a simple \emph{vector} transformation, which means that the vector structure of the equations of motion is preserved by the transformation. As we shall see below, it leads to compact and coordinate-independent expressions for all the coefficients and operators describing magnetization dynamics.

Vector $\vec{s}$ lies in a two-dimensional plane that is naturally embedded in the three-dimensional $(x, y, z)$ space. There are two methods how such vectors can be described in technical calculations. First, one can choose two unit vectors, $\vec{e}_1$ and $\vec{e}_2$, in the SEV plane and describe $\vec{s}$ using coordinates $s_1$ and $s_2$. This method is easy to implement in the case of a uniform ground state ($\vec{m}_0 \ne f(\vec{r})$), when the unit vectors $\vec{e}_1$ and $\vec{e}_2$ may also be chosen independently of $\vec{r}$. The choice of unit vectors $\vec{e}_1(\vec{r})$ and $\vec{e}_2(\vec{r})$, however, is much more complicated in the case of non-uniform state $\vec{m}_0(\vec{r})$ and this problem even may not have a regular solution at all (due to the ``hairy ball'' theorem). In such non-uniformm cases it is much easier to describe $\vec{s}$ as a point in the embedding three-dimensional space subject to the orthogonality constraint $\vec{m}_0 \cdot \vec{s} = 0$. This method is analogous to the description of the unit magnetization vector $\vec{m}$ (which belongs to a two-dimensional manifold -- unit sphere) using three Cartesian coordinates $m_x$, $m_y$, and $m_z$ subject to the constraint $m_x^2 + m_y^2 + m_z^2 = 1$. The presented below theory is written in a coordinate-independent vector form, and one can use either method in technical calculations. All main equations of the theory were specifically written in the form in which the orthogonality constraint is automatically satisfied (similar to automatic satisfaction of the condition $d|\vec{m}|/dt = 0$ by the Landau-Lifshits equation). We would also like to note, that in the classical complex Hamiltonian formalism of magnetization dynamics, which uses similar basic ideas, one always has to use the first description method (using unit vectors in the plane), and this is one of the reasons why this method has never been successfully used for analysis of spin excitations on a non-uniform background.

The transformation Eq.~(\ref{mapping}) maps the whole unit sphere into the disk $|\vec{s}| < 2$ (note that the unit sphere and the disk have the same area of $4\pi$). In the ground state $\vec{m} = \vec{m}_0$ the spin excitation vector is equal to zero, $\vec{s} = 0$, thus the SEV $\vec{s}$ is a measure of deviation of magnetization from the ground state (measure of excitation of the spin system). The mapping of the antipode point $\vec{m} = -\vec{m}_0$ is not uniquely determined by Eq.~(\ref{mapping}) (it is mapped into the whole circle $|\vec{s}| = 2$), which is connected with different topologies of a sphere and a plane. In reality, Eq.~(\ref{mapping}) is useful for description of weakly nonlinear magnetization dynamics $|\vec{s}| \ll 2$. Note, however, that the ground state $\vec{m}_0$ itself can be strongy non-uniform, \ie the proposed approach can be used to describe weakly nonlinear dynamics of such magnetic objects as domain walls, vortices, or skyrmions. In the weakly nonlinear case $|\vec{s}| \ll 2$ Eq.~(\ref{mapping}) can be expanded in the Taylor series as
\begin{equation}\label{mapping-Taylor}
	\vec{m} = \vec{m}_0 + \vec{s} - \frac{s^2}{2} \, \vec{m}_0 - \frac{s^2}{8} \, \vec{s} + O(s^5)
\,.\end{equation}
As one can see, in the linear limit the SEV $\vec{s}$ is equal to the magnetization deviation $(\vec{m} - \vec{m}_0)$, which makes it a very convenient object for mixing analytical and numerical approaches: practically at any point of the theoretical analysis one can use interchangeably either of these methods to solve a particular part of the problem, and switching from one approach to another does not require any additional transformations.

The inverse to Eq.~(\ref{mapping}) transformation is given by
\begin{eqnarray}\label{mapping-inverse}
	\vec{s} &=& \sqrt{\frac{2}{1 + \vec{m}_0 \cdot \vec{m}}} \, \op{P}_0 \cdot \vec{m}
		\\\nonumber
		&=& \op{P}_0 \cdot \vec{m} + \frac{1 - \vec{m}_0 \cdot \vec{m}}{4} \, \op{P}_0 \cdot \vec{m} + O(|\vec{m} - \vec{m}_0|^5)
\,,\end{eqnarray}
where $\op{P}_0$ is the projection operator into the SEV plane (plane orthogonal to $\vec{m}_0$):
\begin{equation}\label{def-P0}
	\op{P}_0 \equiv \op{I} - \vec{m}_0 \otimes \vec{m}_0
\,.\end{equation}

In terms of the spin excitation vector $\vec{s}$, the area element $d\A$ Eq.~(\ref{def-dA}) has a simple form
\begin{equation}\label{dA-s}
	d\A = \frac12 \, \vec{m}_0 \cdot (\vec{s} \times d\vec{s})
\,,\end{equation}
and the magnetic Lagrangian Eq.~(\ref{LLE-Lagrangian}) can be written as
\begin{equation}\label{Lagrangian-s}
	\L = \frac12 \int_{V_s} \vec{s} \cdot \op{L}_0 \cdot \frac{d\vec{s}}{dt} \, d\vec{r} - \H(\vec{s})
\,,\end{equation}
where the skew-symmetric operator $\op{L}_0$ is defined by
\begin{equation}\label{def-opL}
	\op{L}_0 \cdot \vec{v} \equiv -L_s \, \vec{m}_0 \times \vec{v}
\,.\end{equation}
Respectively, the equation of motion for the SEV $\vec{s}$ has the form of a vector Hamiltonian equation
\begin{equation}\label{eqmot-s}
	\op{L}_0 \cdot \frac{d\vec{s}}{dt} = \frac{\delta\H}{\delta\vec{s}}
\,.\end{equation}

Note, that the operator $\op{L}_0$ is invertible for vectors $\vec{s}$ orthogonal to $\vec{m}_0$:
\begin{equation}\nonumber
	\op{L}_0^2 \cdot \vec{s} = -L_s^2 \vec{s}
\,,\end{equation}
and Eq.~(\ref{eqmot-s}) can also be written as
\begin{equation}\nonumber
	\frac{d\vec{s}}{dt} = -L_s^{-2} \op{L}_0 \cdot \frac{\delta\H}{\delta\vec{s}}
\,.\end{equation}
Thus, Eq.~(\ref{eqmot-s}) is a well-defined dynamical equation for the SEV $\vec{s}$.

The Hamiltonian $\H(\vec{s})$ of a magnetic system is obtained by substituting the transformation Eq.~(\ref{mapping}) into Eq.~(\ref{energy-1}). Weakly-nonlinear expansion of $\H(\vec{s})$ reads
\begin{equation}\label{energy-s}
	\H = \H_2 + \H_3 + \H_4
\,,\end{equation}
where
\begin{subequations}\label{energy-terms}
\begin{eqnarray}\label{energy-terms-2}
	\H_2 &=& \frac12 \int_{V_s} \vec{s} \cdot \op{H}_0 \cdot \vec{s} \, d\vec{r}
		\,,\\\label{energy-terms-3}
	\H_3 &=& -\frac12 \int_{V_s} (s^2 \vec{m}_0) \cdot \op{H} \cdot \vec{s} \, d\vec{r}
		\,,\\\label{energy-terms-4}
	\H_4 &=& \frac18 \int_{V_s} \Big[
			(s^2 \vec{m}_0) \cdot \op{H} \cdot (s^2 \vec{m}_0)
		\\\nonumber&&
			\qquad\qquad - (s^2 \vec{s}) \cdot \op{H} \cdot \vec{s}
		\Big] \, d\vec{r}
\,,\end{eqnarray}
\end{subequations}
and the linear Hamiltonian of the system $\op{H}_0$ is defined as
\begin{equation}\label{def-opH0}
	\op{H}_0 \equiv \op{P}_0 \cdot (\op{H} + M_s B_0 \op{I}) \cdot \op{P}_0
\,.\end{equation}
We kept in Eq.~(\ref{energy-s}) only the terms up to fourth order in $|\vec{s}|$, which is sufficient for most weakly nonlinear problems. We have also added $\op{P}_0$ projectors in the definition of the operator $\op{H}_0$ Eq.~(\ref{def-opH0}). This does not change the quadratic part of the energy $\H_2$ Eq.~(\ref{energy-terms-2}) since $\op{P}_0 \cdot \vec{s} = \vec{s}$, but it is convenient for further analysis because now one can find the variational derivative $\delta\H_2/\delta\vec{s} = \op{H}_0 \cdot \vec{s}$ without explicit taking into account the orthogonality condition $\vec{m}_0 \cdot \vec{s} = 0$. Note, also, that the operator $\op{H}_0$ defined by Eq.~(\ref{def-opH0}) is a self-adjoint operator, which will be important in the following.

It is also interesting to note, that the nonlinear energy terms $\H_3$ and $\H_4$ Eqs.~(\ref{energy-terms-3})-(\ref{energy-terms-4}) do not explicitly depend on the internal magnetic field $B_0$, and any dependence of nonlinear properties of spin excitations on magnetic field can be explained by the field dependence of the profiles of spin wave modes and field dependence of the ground state $\vec{m}_0$.

Introduction of the spin excitation vector $\vec{s}$ allowed us to formulate the magnetization dynamics in a ``flat'' phase space, to which standard methods of weakly nonlinear dynamical systems can be directly applied. Namely, weakly-nonlinear dynamics of a magnetic system is most easily described in terms of amplitudes of linear spin wave modes. The technical details of this approach are derived in the rest of this Section.

\subsection{Linear Eigenmodes of a Magnetic System}
\label{ss:Eigenmodes}

In the limit of linear excitations Eq.~(\ref{eqmot-s}) becomes
\begin{equation}\label{eqmot-s-linear}
	\op{L}_0 \cdot \frac{d\vec{s}}{dt} = \op{H}_0 \cdot \vec{s}
\,.\end{equation}
The harmonic solutions ($d/dt \rightarrow -i\omega_\alpha$) of this equation $\vec{s}_\alpha$ are the linear eigenmodes of magnetic excitations:
\begin{equation}\label{eigenproblem}
	-i\omega_\alpha \op{L}_0 \cdot \vec{s}_\alpha = \op{H}_0 \cdot \vec{s}_\alpha
\,.\end{equation}
Here $\alpha$ is the mode index, $\omega_\alpha$ is its eigenfrequency, and $\vec{s}_\alpha \equiv \vec{s}_\alpha(\vec{r})$ is the complex eigenmode profile.

In Eq.~(\ref{eigenproblem}) $\op{L}_0$ is a skew-symmetric operator, while $\op{H}_0$ is a symmetric (Hermitian) operator. Thus, this equation is a generalized Hamiltonian eigenvalue problem, properties of which are well studied. In an important case when the operator $\op{H}_0$ is positive-definite (\ie the magnetic ground state $\vec{m}_0$ corresponds to a minimum of energy) the eigenvectors $\vec{s}_\alpha$ form a complete set (basis) in the space of vector functions orthogonal to the ground magnetic state $\vec{m}_0$, and all the eigenfrequencies $\omega_\alpha$ are real-valued (see Appendix~\ref{app:Eigenproblem}). This may also be true in a case when $\op{H}_0$ is not positive-definite, although there is no guarantee. In the following, we shall assume that this important property holds.

Using Eq.~(\ref{eigenproblem}) and symmetry properties of the operators $\op{L}_0$ and $\op{H}_0$ one can derive two orthogonality relations for the mode profiles $\vec{s}_\alpha(\vec{r})$ (see Appendix~\ref{app:Eigenproblem} for mathematical details). The first relation is
\begin{equation}\label{orthogonality}
	\int_{V_s} \vec{s}_\alpha^* \cdot \op{L}_0 \cdot \vec{s}_{\alpha'} \, d\vec{r} = i \hbar_\alpha \Delta_{\alpha, \alpha'}
\,,\end{equation}
where $\Delta_{\alpha, \alpha'}$ is the Kronecker delta and $\hbar_\alpha$ is the real-valued norm of the $\alpha$-th mode:
\begin{equation}\label{def-norm}
	\hbar_\alpha \equiv -i \int_{V_s} \vec{s}_\alpha^* \cdot \op{L}_0 \cdot \vec{s}_\alpha \, d\vec{r}
\,.\end{equation}
In case of degenerate spectrum (several modes have the same frequency $\omega_\alpha$) the relation Eq.~(\ref{orthogonality}) should be understood in the usual sense, \ie that it is possible to choose such combinations of degenerate eigenvectors that the relation Eq.~(\ref{orthogonality}) holds.

The norm $\hbar_\alpha$ has dimensionality of action (angular momentum) and is always real-valued and non-zero (but not necessary positive). As it is clear from the definition Eq.~(\ref{def-norm}) and form of the magnetic Lagrangian Eq.~(\ref{Lagrangian-s}), the norm $\hbar_\alpha$ equals the reduced action corresponding to the single-mode excitation and one period of oscillations $2\pi/\omega_\alpha$, divided by $2\pi$. The choice $\hbar_\alpha = \hbar$ (the usual reduced Planck constant) corresponds to the quasi-classical ``magnon'' normalization, when the mode profile $\vec{s}_\alpha$ is a classical analog of a magnon wavefunction and the mode amplitude squared equals the number of magnons in a given quantum state. This analogy explains our choice of notation for the mode norm $\hbar_\alpha$. Another choice of normalization, which may be useful in certain applications, is the normalization to the total spin of the magnetic system, $\hbar_\alpha = L_s V_s$ [for a system with inhomogeneous spin density $L_s(\vec{r})$, $\hbar_\alpha = \int_{V_s} L_s(\vec{r}) \, d\vec{r}$], when the mode amplitude is directly proportional to the magnetization precession angle. We shall not specify a particular choice of normalization and all the expressions presented here are valid for any choice of $\hbar_\alpha$, including the cases when different modes are normalized differently.

The orthogonality relations Eq.~(\ref{orthogonality}) allows one to project an arbitrary vector function into particular eigenstates and are necessary for development of a general perturbation theory (see Sec.~\ref{s:Perturbation}).

Another orthogonality-type relation which follows from Eq.~(\ref{eigenproblem}) has the form
\begin{equation}\label{ortho-energy}
	\int_{V_s} \vec{s}_\alpha^* \cdot \op{H}_0 \cdot \vec{s}_{\alpha'} \, d\vec{r} = \hbar_\alpha \omega_\alpha \Delta_{\alpha,\alpha'}
\,.\end{equation}
This relation can be used for precise (variationally-stable) determination of the eigenfrequencies $\omega_\alpha$ from approximate spatial profiles $\vec{s}_\alpha(\vec{r})$ of the eigenmodes.

Both operators $\op{L}_0$ and $\op{H}_0$ in Eq.~(\ref{eigenproblem}) are real-valued. Then, if $\vec{s}_\alpha(\vec{r})$ is an eigenfunction with eigenvalue $\omega_\alpha$ and norm $\hbar_\alpha$, then the complex-conjugated vector $\vec{s}_\alpha^*(\vec{r})$ is also an eigenfunction corresponding to the eigenvalue $-\omega_\alpha$ and norm $-\hbar_\alpha$. Such ``doubling'' of eigenfunctions is a direct consequence of real-valuedness of the LLE Eq.~(\ref{LLE}) and symmetry of the frequency spectrum of any real process. Thus, only half of the formal eigenmodes of Eq.~(\ref{eigenproblem}) are independent and describe ``physical'' modes; the other half are the formal ``conjugated'' modes that guarantee real-valuedness of the SEV $\vec{s}(t, \vec{r})$. The physical modes are the modes with positive norm $\hbar_\alpha > 0$; as it is clear from Eq.~(\ref{ortho-energy}), such modes correspond to positive eigenvalues $\omega_\alpha > 0$ if the operator $\op{H}_0$ is positive-definite (\ie the ground state $\vec{m}_0$ is an energy minimum). Respectively, conjugated modes have negative norms $\hbar_\alpha < 0$ and, for a positive-definite $\op{H}_0$, negative frequencies $\omega_\alpha < 0$.

In all the above equations the mode index $\alpha$ enumerated all formal modes, both physical and conjugated. We shall keep these notations below and will use indices $\alpha$, $\beta$, $\dots$ to enumerate or sum over all formal modes. To indicate only the physical modes, we shall use indices $\nu$, $\mu$, $\dots$. The notation $\alpha^*$ will be used to indicate a mode ``conjugated'' to the mode $\alpha$, \ie
\begin{equation}\nonumber
	\vec{s}_{\alpha^*} \equiv \vec{s}_\alpha^*
\,,\quad
	\omega_{\alpha^*} \equiv -\omega_\alpha
\,,\quad
	\hbar_{\alpha^*} \equiv -\hbar_\alpha
\,.\end{equation}

Finally, we shall make a note on numerical determination of the eigenfrequencies $\omega_\alpha$ and eigenvectors $\vec{s}_\alpha(\vec{r})$. The technically simplest (although not the best) way to find them, which is often used in modern research, is by using Fourier analysis on results of direct numerical simulations of free-decaying magnetization dynamics after an initial low-amplitude (linear) perturbation from the ground state. In this case the eigenfrequencies $\omega_\alpha$ can be identified from peak positions in the magnetization precession spectrum, and the eigenvectors $\vec{s}_\alpha(\vec{r})$ can be found as cell-by-cell Fourier images of the magnetization at the Fourier frequencies $\omega = \omega_\alpha$. The advantage of the VHF is that the mode profiles found from such numerical procedure \emph{exactly} coincide with the theoretical eigenvectors $\vec{s}_\alpha$, do not require any post-processing (except from possible normalization to provide desired norms $\hbar_\alpha$), and can be directly used in further analysis. This straightforward method is very simple, can be performed using standard micromagnetic packages and processing techniques, and often produce sufficiently accurate results.

The eigenvalue problem Eq.~(\ref{eigenproblem}) can also be solved directly. In most cases, one is interested only in a small fraction of all formally possible magnetic modes, usually the modes with smallest eigenfrequencies $|\omega_\alpha|$. Then, one can use a modification of the Ritz or gradient descent methods based on the variationally-stable expression Eq.~(\ref{ortho-energy}). The derived orthogonality conditions Eq.~(\ref{orthogonality}) and Eq.~(\ref{ortho-energy}) also allow one to adopt Arnoldi iteration and Lanczos algorithm for solution of magnetic eigenvalue problems. These two methods are especially promising for study of magnetic excitations in large magnetic systems since they do not require explicit matrix representation of the energy operator $\op{H}$, but only calculation of action of this operator on individual SEVs $\vec{s}(\vec{r})$, which can be easily accomplished using standard micromagnetic packages with the help of Eq.~(\ref{Beff-2}).

As it was already mentioned above, the SEV $\vec{s}(\vec{r})$ at every point $\vec{r}$ is a two-dimensional vector due to the restriction $\vec{m}_0 \cdot \vec{s} = 0$. In numerical analysis, however, it is much more convenient to describe it as a three-dimensional vector. Such three-dimensional description will, of course, lead to appearance of spurious unphysical formal modes (with $\vec{s}_\alpha || \vec{m}_0$). In the VHF formulation Eq.~(\ref{eigenproblem}), however, these spurious modes do not represent any problem, since all of them correspond to zero eigenfrequency $\omega_\alpha = 0$ and can be automatically filtered out.

The described in this subsection general properties of the linear eigenmodes of magnetic excitations allows one to formulate nonlinear magnetization dynamics using the standard language of complex mode amplitudes and use standard and well-developed Hamiltonian techniques for its analysis.

\subsection{Eigenmode Expansion of the Spin Excitation Vector}
\label{ss:Expansion}

The linear eigenmodes of a magnetic system $\vec{s}_\alpha(\vec{r})$ form a complete set of vector functions orthogonal to the ground state $\vec{m}_0(\vec{r})$. Therefore, any time-dependent spin excitation vector $\vec{s}(t, \vec{r})$ can be expanded in a series over these eigenmodes:
\begin{equation}\label{expansion}
	\vec{s}(t, \vec{r}) = \sum_\alpha \vec{s}_\alpha(\vec{r}) c_\alpha(t)
		= \sum_\nu \left[\vec{s}_\nu(\vec{r}) c_\nu(t) + \cc\right]
\,.\end{equation}
Here $c_\alpha(t)$ is the complex amplitude of the $\alpha$-th mode ($c_{\alpha^*} = c_\alpha^*$) and $\cc$ stands for complex-conjugated part. The first version of the expansion (sum over $\alpha$) uses summation over all formal (``normal'' and ``conjugated'') modes, while the second version (sum over $\nu$) explicitly sums up only the ``normal'' modes, whereas the ``conjugated'' modes are automatically included in the ``$\cc$'' part. The two versions are completely mathematically equivalent, but differ in convenience of use: while the first ($\alpha$) version is more convenient in derivation of general properties and relations, the second one ($\nu$) is preferable in practical analytical or numerical calculations of spin wave dynamics.

Using the orthogonality relations for the eigenmodes $\vec{s}_\alpha$, one can easily reformulate the magnetization dynamics in terms of the mode amplitudes $c_\alpha$. Thus, the Lagrangian Eq.~(\ref{Lagrangian-s}) of the magnetic system takes the form
\begin{equation}\label{Lagrangian-c}
	\L = \frac{i}{2}\sum_\alpha \hbar_\alpha c_{\alpha^*}\frac{dc_\alpha}{dt} - \H
		= \frac{i}{2}\sum_\nu \hbar_\nu \left( c_\nu^*\frac{dc_\nu}{dt} - \cc \right)- \H
\,,\end{equation}
which induces the Hamiltonian equations of motion for the amplitudes $c_\alpha$:
\begin{equation}\label{eqmot-c}
	i \hbar_\alpha \frac{dc_\alpha}{dt} = \frac{\partial\H}{\partial c_{\alpha^*}}
\,.\end{equation}
The dynamical equation written in this form is valid for both normal and conjugated modes (due to the property $\hbar_{\alpha^*} = -\hbar_\alpha$). It can also be rewritten as
\begin{equation}\label{eqmot-Poisson}
	\frac{dc_\alpha}{dt} = [\H, c_\alpha]
\,,\end{equation}
where $[\cdot, \cdot]$ is the Poisson brackets of the system:
\begin{equation}\label{def-Poisson}
	[A, B] \equiv \sum_\alpha \frac{i}{\hbar_\alpha} 
		\frac{\partial A}{\partial c_\alpha} \frac{\partial B}{\partial c_{\alpha^*}}
\,.\end{equation}
The Poisson-bracket form of the equations of motion can also be written for any function $F = F(t, \{c_\alpha\})$ of time and complex spin wave amplitudes $c_\alpha$:
\begin{equation}\label{eqmot-Poisson-f}
	\frac{dF}{dt} = [\H, F] + \frac{\partial F}{\partial t}
\,.\end{equation}

The weakly-nonlinear expansion of the Hamiltonian $\H$ has the form Eq.~(\ref{energy-s}), where different-order terms are expressed through the amplitudes $c_\alpha$ as
\begin{subequations}\label{energy-terms-c}
\begin{eqnarray}\label{energy-terms-c-2}
	\H_2 &=& \frac12\sum_\alpha \hbar_\alpha \omega_\alpha |c_\alpha|^2
		= \sum_\nu \hbar_\nu \omega_\nu |c_\nu|^2
\,,\\\label{energy-terms-c-3}
	\H_3 &=& \frac16 \sum_{\alpha\beta\gamma} V_{\alpha\beta\gamma} 
		c_{\alpha} c_{\beta} c_{\gamma}
\,,\\\label{energy-terms-c-4}
	\H_4 &=& \frac{1}{24} \sum_{\alpha\beta\gamma\delta} W_{\alpha\beta\gamma\delta} 
		c_{\alpha} c_{\beta} c_{\gamma} c_{\delta}
\,.\end{eqnarray}
\end{subequations}
Here
\begin{subequations}\label{et-comp}
\begin{eqnarray}
	V_{\alpha\beta\gamma} &=& \tilde{V}_{\alpha\beta,\gamma}
		+ \tilde{V}_{\beta\gamma,\alpha} + \tilde{V}_{\gamma\alpha,\beta}
\,,\\
	\tilde{V}_{\alpha\beta,\gamma} &=& -\int_{V_s} ((\vec{s}_\alpha \cdot \vec{s}_\beta) \vec{m}_0) \cdot \op{H} \cdot \vec{s}_\gamma \, d\vec{r}
\,,\\
	W_{\alpha\beta\gamma\delta} &=& \tilde{W}_{\alpha\beta,\gamma\delta}
		+ \tilde{W}_{\alpha\gamma,\beta\delta} + W_{\alpha\delta,\beta\gamma}
\,,\\
	\tilde{W}_{\alpha\beta,\gamma\delta} &=& \int_{V_s} \Bigg[
			((\vec{s}_\alpha \cdot \vec{s}_\beta) \vec{m}_0) \cdot \op{H} \cdot ((\vec{s}_\gamma \cdot \vec{s}_\delta) \vec{m}_0)
		\\\nonumber&&\hspace*{-3.5em}
			-\frac14 ((\vec{s}_\alpha \cdot \vec{s}_\beta) \vec{s}_\gamma) \cdot \op{H} \cdot \vec{s}_\delta
			-\frac14 ((\vec{s}_\alpha \cdot \vec{s}_\beta) \vec{s}_\delta) \cdot \op{H} \cdot \vec{s}_\gamma
		\\\nonumber&&\hspace*{-3.5em}
			-\frac14 ((\vec{s}_\gamma \cdot \vec{s}_\delta) \vec{s}_\alpha) \cdot \op{H} \cdot \vec{s}_\beta
			-\frac14 ((\vec{s}_\gamma \cdot \vec{s}_\delta) \vec{s}_\beta) \cdot \op{H} \cdot \vec{s}_\alpha
		\Bigg] \, d\vec{r}
\,.\end{eqnarray}
\end{subequations}
These expressions for the tree-magnon $V_{\alpha\beta\gamma}$ and four-magnon $W_{\alpha\beta\gamma\delta}$ interaction coefficients look complicated, however, their calculation requires only evaluation of various ``matrix elements'' of the energy operator $\op{H}$ with various combinations of the eigenmode profiles $\vec{s}_\alpha(\vec{r})$. Calculation of such ``matrix elements'' can be easily done numerically once the mode profiles are found either analytically or from micromagnetic simulations. The interaction coefficients $V_{\alpha\beta\gamma}$ and $W_{\alpha\beta\gamma\delta}$ were symmetrized with respect to exchange of any pair of indices (\eg $V_{\alpha\beta\gamma} = V_{\beta\alpha\gamma} = V_{\alpha\gamma\beta}$), which is the reason for many combinatorial terms in the definition Eqs.~(\ref{et-comp}) and complicated form of these equations.

According to our convention, the nonlinear energy terms in Eqs.~(\ref{energy-terms-c}) are written using summation over all formal modes (both ``normal'' $\nu$ and ``conjugated'' $\nu^*$), which leads to more compact form of these terms. However, one has to remember this fact when interpreting possible nonlinear processes described by Hamiltonian of different orders. For instance, the three-wave Hamiltonian $\H_3$ contains terms proportional to $c_{\nu}^* c_{\nu'}^* c_{\nu''}$ ($\alpha = \nu^*$, $\beta = (\nu')^*$, $\gamma = \nu''$), which, in analogy with quantum physics, can be interpreted as a process of parametric decay of magnon $\nu''$ into two magnons $\nu$ and $\nu'$, and terms proportional to $c_{\nu}^* c_{\nu'}^* c_{\nu''}^*$ ($\alpha = \nu^*$, $\beta = (\nu')^*$, $\gamma = (\nu'')^*$), which describe creation of three magnons from vacuum state. The Hamiltonian $\H_3$ also contains ``conjugated'' processes (proportional to $c_{\nu} c_{\nu'}^* c_{\nu''}^*$ and $c_{\nu} c_{\nu'} c_{\nu''}$), and the interaction coefficients describing direct and conjugated processes are complex conjugates of each other:
\begin{equation} 
	V_{\alpha^*\beta^*\gamma^*} = V_{\alpha\beta\gamma}^*
\,,\qquad
	W_{\alpha^*\beta^*\gamma^*\delta^*} = W_{\alpha\beta\gamma\delta}^*
\,.\end{equation}

In many cases, the nonlinear processes describing three-wave interactions are non-resonant in the sense that
\begin{equation}\label{nonresonant-3}
	\omega_{\alpha\beta\gamma} \equiv \omega_\alpha + \omega_\beta + \omega_\gamma \ne 0
\end{equation}
for any set of modes for which $V_{\alpha\beta\gamma} \ne 0$. This is, obviously, always the case for the processes of three-magnon creation from vacuum $c_{\nu}^* c_{\nu'}^* c_{\nu''}^*$ (if the ground state corresponds to a minimum of energy), but may also be true for parametric decay processes $c_{\nu}^* c_{\nu'}^* c_{\nu''}$. Such non-resonant three-magnon processes can be eliminated by a weakly-nonlinear canonical transformation of complex spin wave amplitudes $c_\alpha$ (see Appendix~\ref{app:Elimination}), leading to a new Hamiltonian with $\H_3 = 0$ and modified four-magnon interaction terms
\begin{subequations}\label{W-renormalized}
\begin{eqnarray}
	W_{\alpha\beta\gamma\delta}' &=& W_{\alpha\beta\gamma\delta}
\\\nonumber&&
		+ \Delta W_{\alpha\beta,\gamma\delta}
		+ \Delta W_{\alpha\gamma,\beta\delta}
		+ \Delta W_{\alpha\delta,\beta\gamma}
\,,\\
	\Delta W_{\alpha\beta,\gamma\delta} &=&
		\sum_\epsilon
			\frac{V_{\alpha\beta\epsilon^*}V_{\epsilon\gamma\delta}}{2\hbar_\epsilon} 
		\left(
			\frac{1}{\omega_{\alpha\beta\epsilon^*}} - \frac{1}{\omega_{\epsilon\gamma\delta}}
		\right)
\,.\end{eqnarray}
\end{subequations}

The four-magnon processes are always resonant, and, therefore, cannot be eliminated by a similar renormalization procedure. For this reason, in most cases it is enough to take into account only three-wave $\H_3$ and four-wave $\H_4$ terms to describe dynamics of any system of weakly-nonlinear excitations.

Using the form of the Hamiltonian Eq.~(\ref{energy-terms-c}), the equation of motion for mode amplitudes $c_\alpha$ Eq.~(\ref{eqmot-c}) can be written explicitly as
\begin{eqnarray}\label{eqmot-c-explicit}
i\hbar_\alpha\frac{dc_\alpha}{dt} &=& \hbar_\alpha \omega_\alpha c_\alpha
	+ \frac12\sum_{\beta\gamma} V_{\alpha^*\beta\gamma} c_{\beta} c_{\gamma}
\\\nonumber&&\qquad
	+ \frac16\sum_{\beta\gamma\delta} W_{\alpha^*\beta\gamma\delta} c_{\beta} c_{\gamma} c_{\delta}
\,.\end{eqnarray}
This equation describes weakly-nonlinear magnetization dynamics in an arbitrary magnetic system. In most practically interesting cases, the number of efficiently (resonantly) interacting spin wave modes is limited and it is enough to take into account only few spin wave modes relevant in a studied nonlinear process. Thus, the transformation of the original Landau-Lifshits equation to a system Eq.~(\ref{eqmot-c-explicit}) usually allows one to substantially reduce the dimensionality of the phase space of the studied system and often enables analytical analysis of rather complicated nonlinear spin wave processes.

\section{Perturbation Theory}
\label{s:Perturbation}

In the previous section we derived Hamiltonian equations of motion for spin wave amplitudes $c_\alpha$ in the case of a conservative magnetic system with time-independent magnetic field. Here we consider modifications of the equations of motion caused by other magnetic interactions, which may be treated perturbatively. There are two different classes of magnetic perturbations, which we shall consider separately.

The first class is the conservative perturbations, which may be described by an additional term $\Delta\H(t, \vec{m})$ in the Hamiltonian of the system. The most important example of conservative perturbations is the interaction of a magnetic system with microwave magnetic field $\vec{b}(t, \vec{r})$, which describes excitation of spin waves by an external system. In this case the perturbation Hamiltonian has the form
\begin{equation}\label{dH-1}
	\Delta\H = -\int M_s \vec{b} \cdot \vec m d\vec{r}
\,.\end{equation}

The second type of perturbations is the non-conservative perturbations, which may be described by the additional torque $\Delta\vec{T}(t, \vec{r}, \vec{m})$ in the right-hand side of LLE Eq.~(\ref{LLE}). The most important example of non-conservative perturbations is the dissipation of spin waves, which may be described by the Gilbert damping torque
\begin{equation}\label{Gilbert-1}
	\Delta\vec{T} = \alpha_G \vec{m} \times \frac{\partial\vec{m}}{\partial t}
\,,\end{equation}
where $\alpha_G$ is the dimensionless Gilbert damping parameter.

Below we shall consider these two examples.

\label{ss:conservative-perturbations}

The spin wave amplitudes $c_\alpha$ of the VHF are Hamiltonian dynamical variables, described by the same Hamiltonian as the original magnetic system. Therefore, analysis of influence of any conservative perturbation within the VHF is very simple: the perturbation will lead to an additional Hamiltonian term in the equation of motion
\begin{equation}\label{add-eqmot-1}
	i\hbar_\alpha\left(\frac{dc_\alpha}{dt}\right)_{\rm pert} = \frac{\partial\Delta\H}{\partial c_{\alpha^*}}
\,,\end{equation}
where the perturbation Hamiltonian $\Delta\H$ should be expressed through the spin wave amplitudes $c_\alpha$.

We shall consider the particular example Eq.~(\ref{dH-1}) of external magnetic field. Substituting into Eq.~(\ref{dH-1}) the approximate expression Eq.~(\ref{mapping-Taylor}) and expanding the SEV $\vec{s}(t, \vec{r})$ over the spin wave modes Eq.~(\ref{expansion}), one obtains the explicit expression for $\Delta\H(t, c_\alpha)$:
\begin{equation}\label{dH-expansion}
	\Delta\H = \Delta\H_1 + \Delta\H_2 + \Delta\H_3
\,,\end{equation}
where we have dropped irrelevant constant energy term and
\begin{subequations}\label{dH-terms}
\begin{eqnarray}
	\Delta\H_1 &=& \sum_\alpha P_\alpha c_\alpha
\,,\\
	\Delta\H_2 &=& \frac12 \sum_{\alpha_1, \alpha_2} Q_{\alpha_1\alpha_2} c_{\alpha_1} c_{\alpha_2}
\,,\\
	\Delta\H_3 &=& \frac16 \sum_{\alpha_1, \alpha_2, \alpha_3} R_{\alpha_1\alpha_2\alpha_3} c_{\alpha_1} c_{\alpha_2} c_{\alpha_3}
\,.\end{eqnarray}
\end{subequations}
Here the excitation coefficients $P_\alpha(t)$, $Q_{\alpha_1\alpha_2}(t)$, and $R_{\alpha_1\alpha_2\alpha_3}(t)$ are given by
\begin{subequations}\label{PQR}
\begin{eqnarray}
	P_\alpha &=& - \int M_s \vec{b} \cdot \vec{s}_\alpha d\vec{r}
\,,\\
	Q_{\alpha_1\alpha_2} &=& \int M_s (\vec{b} \cdot \vec{m}_0) (\vec{s}_{\alpha_1} \cdot \vec{s}_{\alpha_2}) d\vec{r}
\,,\\
	R_{\alpha_1\alpha_2\alpha_3} &=& \tilde{R}_{\alpha_1, \alpha_2\alpha_3} + \tilde{R}_{\alpha_2, \alpha_3\alpha_1} 
		+ \tilde{R}_{\alpha_3, \alpha_1\alpha_2}
\,,\\
	\tilde{R}_{\alpha_1, \alpha_2\alpha_3} &=& \frac14 \int M_s (\vec{b} \cdot \vec{s}_{\alpha_1}) (\vec{s}_{\alpha_2} \cdot \vec{s}_{\alpha_3}) d\vec{r}
\,.\end{eqnarray}
\end{subequations}

Similarly to the nonlinear self-interaction coefficients $V_{\alpha_1\alpha_2\alpha_3}$ and $W_{\alpha_1\alpha_2\alpha_3\alpha_4}$, the coefficients $P_\alpha$, $Q_{\alpha_1\alpha_2}$, and $R_{\alpha_1\alpha_2\alpha_3}$, which describe interaction of the spin system with external field, are expressed as simple ``overlap integrals'' and their calculation does not represent any difficulty once the spin wave profiles $\vec{s}_\alpha(\vec{r})$ are known.

Finally, we can write explicit expression for the excitation term in the equation of motion for spin wave amplitudes $c_\alpha$:
\begin{equation}\label{add-eqmot-2}
	i\hbar_\alpha\left(\frac{dc_\alpha}{dt}\right)_{\rm excitation} =
		P_{\alpha^*} + \sum_{\alpha_1} Q_{\alpha^*\alpha_1} c_{\alpha_1} 
		+ \frac12 \sum_{\alpha_1\alpha_2} R_{\alpha^*\alpha_1\alpha_2} c_{\alpha_1} c_{\alpha_2}
\,.\end{equation}
The first term in the right-hand side of this equation ($P_{\alpha^*}$) describes linear excitation of spin waves, the second term ($Q_{\alpha^*\alpha_1}$) -- parametric processes (second-order Suhl processes) and shift of spin wave frequencies due to magnetic field modulation, and the last term ($R_{\alpha^*\alpha_1\alpha_2}$) describes nonlinear corrections to the efficiency of spin wave excitation.

\subsection{Non-Conservative Perturbations -- Gilbert Damping}
\label{ss:non-conservative-perturbations}

In the case of non-conservative perturbations $(\partial\vec{m}/\partial t)_{\rm nc} = \Delta\vec{T}(t, \vec{r}, \vec{m})$ the additional terms in the equation of motion for SEV $\vec{s}(t, \vec{r})$ can be obtained by differentiating approximate Eq.~(\ref{mapping-inverse}) with respect to time $t$:
\begin{equation}\label{nc-eqmot-1}
	\left(\frac{\partial\vec{s}}{\partial t}\right)_{\rm nc} = \op{P}_0 \cdot \Delta\vec{T} 
		- \frac14\left(1 - \frac{s^2}{8}\right) \vec{s} \left(\vec{m}_0 \cdot \Delta\vec{T}\right)
		+ \frac{s^2}{8} \op{P}_0 \cdot \Delta\vec{T}
\,,\end{equation}
where $\Delta\vec{T}$ should be written as a function of $\vec{s}$ using Eq.~(\ref{mapping-Taylor}).

To obtain equations for spin wave amplitudes $c_\alpha$, one should substitute into Eq.~(\ref{nc-eqmot-1}) the expansion Eq.~(\ref{expansion}) and apply orthogonality relation Eq.~(\ref{orthogonality}) by taking scalar product of this equation with $\vec{s}_\alpha^* \cdot \op{L}_0$ and integrating over the volume of the magnetic body. 

Previously, this technique has been used to evaluate damping rates of inhomogeneous spin wave modes caused by various dissipation mechanisms \cite{Verba2018}. For completeness of the VHF presentation, we shall briefly repeat the derivation here. For simplicity, we shall consider only the Gilbert damping mechanism and dissipation only in the linear approximation (in any case, the dissipative torques are phenomenological, and phenomenological nonlinear damping corrections can be added later directly into equations for $c_\alpha$). In the limit of linear deviations, Eq.~(\ref{nc-eqmot-1}) simplifies to the trivial expression
\begin{equation}\label{nc-eqmot-2}
	\left(\frac{\partial\vec{s}}{\partial t}\right)_{\rm nc} = \Delta\vec{T} 
\,.\end{equation}
In this approximation, the Gilbert torque Eq.~(\ref{Gilbert-1}) can be written as
\begin{equation}\label{Gilbert-2}
	\Delta\vec{T} = \alpha_G \vec{m}_0 \times \frac{\partial\vec{s}}{\partial t}
\,.\end{equation}

Using the expansion Eq.~(\ref{expansion}) in Eqs.~(\ref{nc-eqmot-2}) and (\ref{Gilbert-2}) gives
\begin{equation}
	\sum_\alpha \vec{s}_\alpha \left(\frac{c_\alpha}{dt}\right)_{\rm nc}
		= \alpha_G \sum_\beta \vec{m}_0 \times \vec{s}_\beta \frac{dc_\beta}{dt}
\,.\end{equation}
With good accuracy, the time derivative $dc_\beta/dt$ in the right-hand side of this equation can be replaced with its linear conservative value $-i\omega_\beta c_\beta$. Then, multiplying this equation by $\vec{s}_\alpha^* \cdot \op{L}_0$ and integrating over the volume of the magnetic system yields
\begin{equation}
	i\hbar_\alpha \left(\frac{c_\alpha}{dt}\right)_{\rm nc} = 
		-i\alpha_G \sum_\beta \omega_\beta \left[\int L_s \vec{s}_\alpha^* \cdot \vec{s}_\beta d\vec{r} \right] c_\beta
\,.\end{equation}
In the case of small damping $\alpha_G$ and non-degenerate spectrum, one can keep only the diagonal term ($\beta = \alpha$) in the sum in the right-hand side of this equation. Then, the dissipative correction to the equation of spin wave amplitude $c_\alpha$ takes the standard form
\begin{equation}
	i\hbar_\alpha \left(\frac{c_\alpha}{dt}\right)_{\rm nc} = -i\hbar_\alpha \Gamma_\alpha c_\alpha
\,,\end{equation}
where $\Gamma_\alpha$ is the damping rate of $\alpha$-th spin wave mode:
\begin{equation}\label{Gamma}
	\Gamma_\alpha = \frac{\alpha_G\omega_\alpha}{\hbar_\alpha} \int L_s|\vec{s}_\alpha|^2 d\vec{r}
\,.\end{equation}
Equation~(\ref{Gamma}), which was first derived in \cite{Verba2018}, gives the most general expression for Gilbert damping rate of a spin wave mode. This expression can be used to calculate damping of spin wave modes in magnetic systems with non-uniform ground state, in inhomogeneous systems consisting of several different magnetic materials, can be applied for spin wave modes with non-trivial spatial structure, and so on.

\section{Example Application of VHF}
\label{s:Application}

Here, we shall illustrate the application of the developed vector Hamiltonian formalism using a nano-scale magnetic element as a simple test system. Namely, the magnetic system that we choose for the test procedures is a rectangular prism with dimensions 80~nm~$\times$~40~nm~$\times$~5~nm and material parameters corresponding to Permalloy. All the VHF calculations and numerical simulations were performed in the absence of the external bias magnetic field. In numerical simulations, the magnetic element was discretized with rectangular mesh with cell sizes 5~nm~$\times$~5~nm~$\times$~5~nm (128 cells in total).

\begin{figure}
	\includegraphics[width=0.5\textwidth]{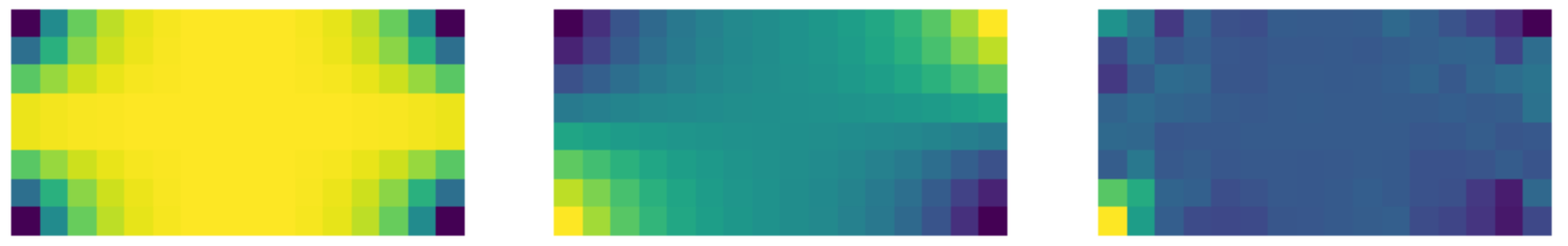}
	\label{f:ground-state}
	\caption{Ground state of the test magnetic system. From left to right: spatial distribution of the equilibrium magnetization components $m_x$, $m_y$, and $m_z$.}
\end{figure}

Figure~\ref{f:ground-state} shows the numerically calculated ground magnetic state of the studied system. It is important to note, that, for such small magnetic prism, the edge effects on magnetization are rather significant and, therefore, the ground state is significantly non-uniform. This means, that the spin wave modes have rather complex profiles that cannot be satisfactory approximated by harmonic functions and, therefore, simple analytical approximations can not be used to describe magnetization dynamics of this system.

\begin{figure}
	\includegraphics[width=0.5\textwidth]{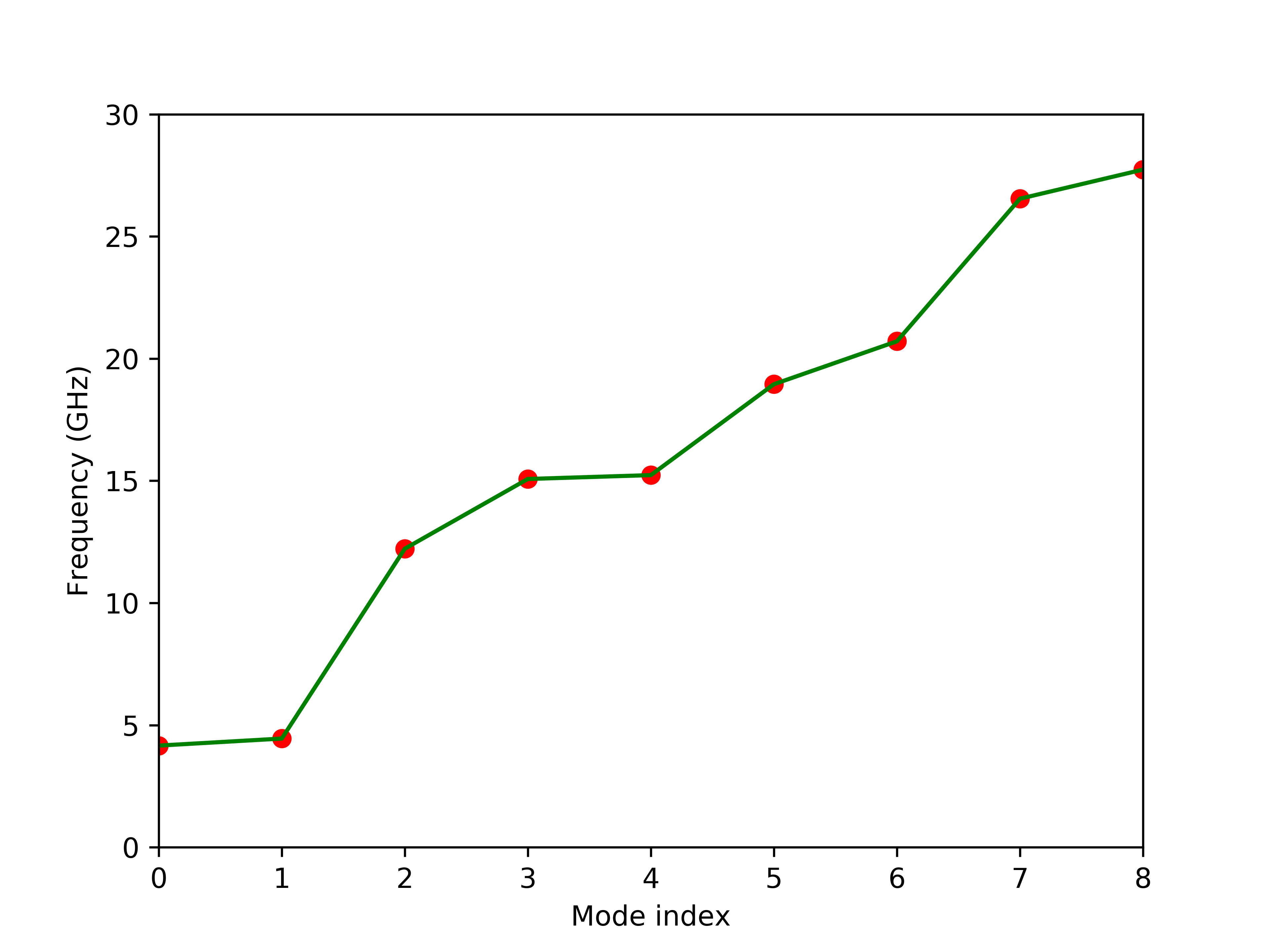}
	\label{f:frequencies}
	\caption{Lowest spin wave eigen-frequencies of the studied magnetic system. Points show eigen-frequencies obtained from direct solution of the linear eigen-value problem Eq.~(\ref{eigenproblem}), solid line -- eigen-frequencies obtained from numerical spin wave profiles using variationally-stable calculation method Eq.~(\ref{ortho-energy}).}
\end{figure}

At the first step we calculated spin wave mode profiles and frequencies of the studied system. Fig.~\ref{f:frequencies} shows dependence of several lowest spin wave eigen-frequencies on mode index. Points show eigen-frequencies obtained from direct numerical solution of the discrete version of the linear eigen-value problem Eq.~(\ref{eigenproblem}), while solid line corresponds to eigen-frequencies obtained from numerical spin wave profiles using variationally-stable calculation method Eq.~(\ref{ortho-energy}). As one can see, spin wave frequencies calculated using these two methods coincide with high precision. This proves validity of the analytical approach for a magnetic system with spatially-nonuniform ground state. Also, this result demonstrates that one can use variationally-stable calculations in the case when the spin wave mode profiles are known only approximately, which may be important for simulations of macro-sized magnetic systems, for which direct solution of the linear eigen-mode problem is not possible and one has to use certain approximate methods.

\begin{figure}
	\includegraphics[width=0.5\textwidth]{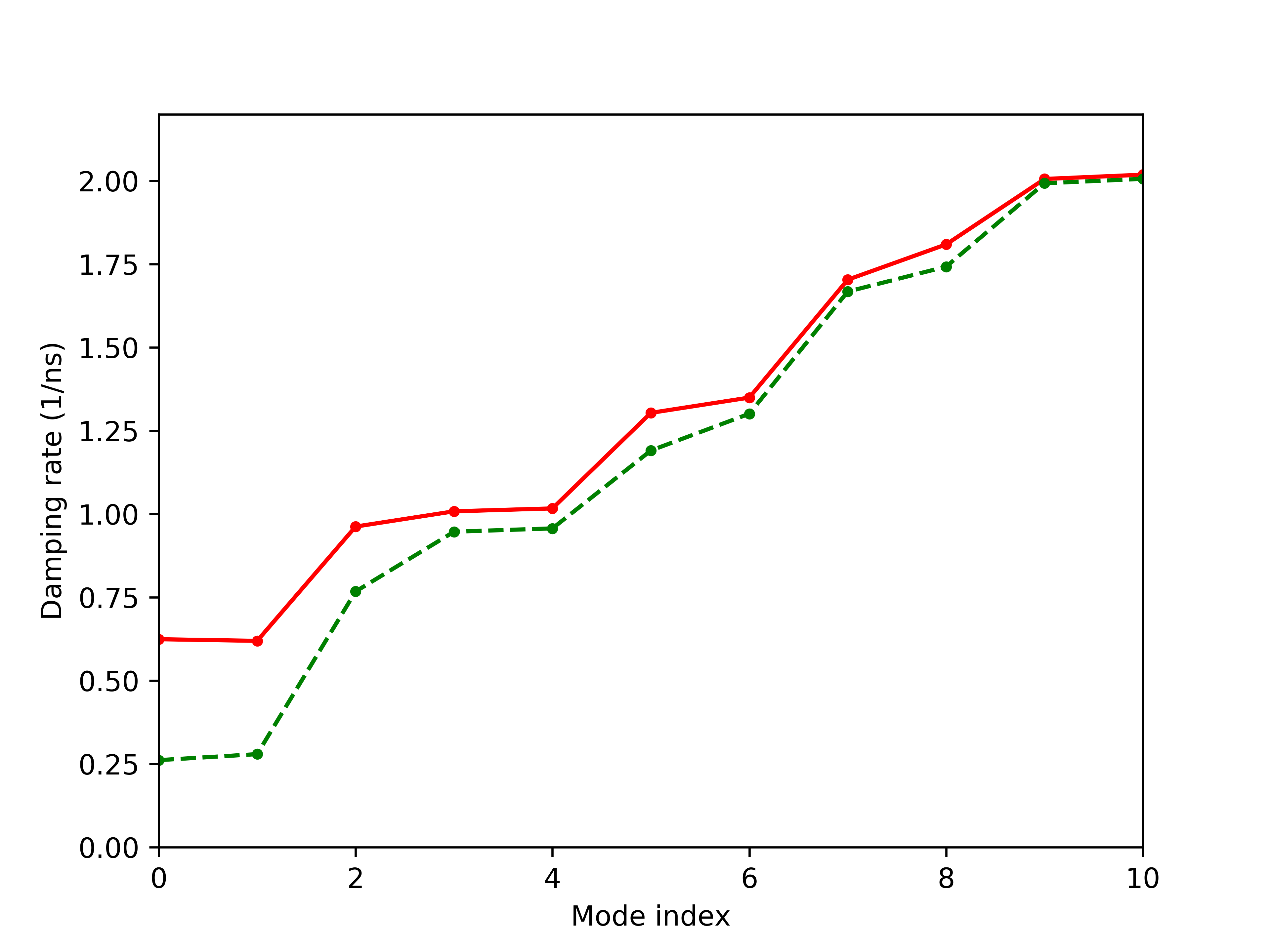}
	\label{f:damping}
	\caption{Damping rate of the lowest spin wave modes of the test magnetic system. Red points and solid line -- damping rate with account of real spatial profile of the spin wave mode Eq.~(\ref{Gamma}); green points and dashed line -- naive approximation $\Gamma = \alpha_G \omega$.}
\end{figure}

Next, we calculated linear damping rate Eq.~(\ref{Gamma}) for all spin wave modes. The results of this calculation are illustrated by Fig.~\ref{f:damping}. Red points joined by the solid line show the result of the VHF calculations, while green points joined by the dashed line correspond to the naive approximation $\Gamma = \alpha_G \omega$, where $\alpha_G=0.01$ is the Gilbert damping constant for Py. As one can see, the two methods give approximately the same damping rates for higher-order spin wave modes, but differ by about a factor of 2 for the spin wave modes with lowest frequencies. This discrepancy is connected with the fact, that the naive Gilbert approximation does not take into account non-uniform profile and ellipticity of precession of spin wave modes. The influence of these factors increase with the decrease of the spin wave frequency. It should be noted, that the modes which are most important from the practical point of view are exactly the lowest-lying spin wave modes, which have non-zero overlap with quasi-uniform magnetic field and, therefore, can be directly excited by an external electromagnetic system. Thus, Fig.~\ref{f:damping} demonstrates that there is a huge difference between damping rates of practically interesting modes calculated using the developed VHF approach and obtained from naive estimations.

We have also calculated various nonlinear interaction coefficients and coefficients of interaction with external microwave field. These parameters will be used below to compare results of VHF analysis with direct numerical simulations for two cases -- linear free magnetization decay and nonlinear ferromagnetic resonance. We would like to stress, that VHF analysis does not have a single fitting parameter and that all coefficients in the VHF equations are calculated as various linear or nonlinear ``overlap integrals'' over the spin wave profiles $\vec{s}_\alpha(\vec{r})$. Calculation of the profiles $\vec{s}_\alpha(\vec{r})$ and eigen-frequencies $\omega_\alpha$ is the only computationally expensive part of the VHF procedure.

\subsection{Linear Free Magnetization Decay}
\label{ss:linear-decay}

To simulate a linear free magnetization decay, we added a small deviation $\vec{\delta m}(\vec{r})$ to the ground magnetic state $\vec{m}_0(\vec{r})$ of the test system and used this non-equilibrium magnetization distribution as the initial condition in full-scale micromagnetic simulations. We run micromagnetic simulations for certain time (approximately 10~ns) and calculated time dependence of $y$ and $z$ components of magnetization, averaged over the volume of the magnetic prism. The same small deviation $\vec{\delta m}(\vec{r})$ was also used as an initial condition for simulations based on the VHF approach. In this case, we projected the initial magnetization deviation into the set of spin wave modes $\vec{s}_\alpha(\vec{r})$, which gave us the initial complex amplitudes of the spin wave modes, $c_\alpha(0)$. Then, we run linear VHF solver to find the time dependence of the complex amplitudes (this dependence is rather trivial in the VHF representation, $c_\alpha(t) = c_\alpha(0)e^{-i\omega_\alpha t-\Gamma_\alpha t}$). Using the obtained time dependence of the spin wave amplitudes $c_\alpha(t)$, we restored the time dependence of the spatial profile of the magnetization and found the averaged values of $y$ and $z$ magnetization components. Thus, each numerical experiment provided two independent sets of data for $m_y(t)$ and $m_z(t)$ obtained using two different approaches -- direct numerical simulations and simulations using the VHF approach. Comparison between these sets of data provided information on accuracy of the VHF-based simulations for linear magnetization dynamics.

Before demonstrating examples of the numerical experiment, we would like to comment on the performance of two simulation methods. Full-scale numerical simulations took approximately the same time (about 1 minute on a laptop we used) for each experiment. The simulations based on the VHF approach were much faster (about 2-3 milliseconds). The extremely small simulation time for the VHF approach is explained by the trivial linear dynamics of spin wave modes in the VHF representation. In the case of full nonlinear VHF simulations, the performance gap is smaller, but is still significant. In the VHF approach, the main computational time is spent on the first VHF initialization step, at which spin wave profiles and eigen-frequencies are calculated. For the chosen test system and method of eigen-problem solution, the duration of the initialization step was about 10 seconds. This time is comparable with the full-scale simulation time of one experiment, but it needs to be performed only once per experimental series. Thus, our linear test demonstrated huge improvement of performance of VHF-based micromagnetic solver compared to traditional approach.

\begin{figure}
	\includegraphics[width=0.5\textwidth]{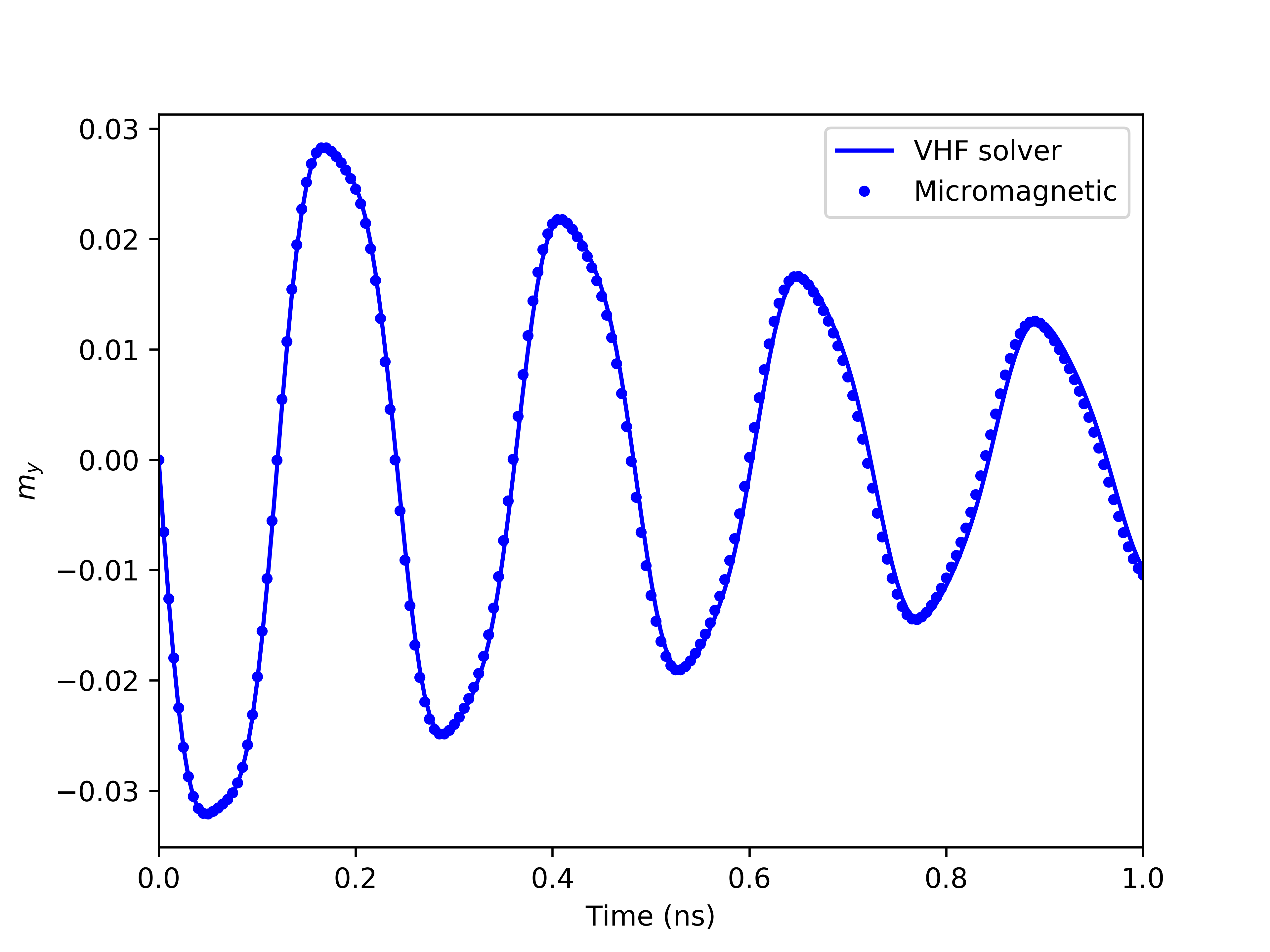}
	\label{f:decay-my}
	\caption{Comparison of time dependence of averaged $m_y$ magnetization component calculated using direct micromagnetic simulations (points) and VHF-based calculations (solid line). Initial magnetization deviation was uniform in space and equal to $\vec{\delta m} = 0.01\vec{y}$.}
\end{figure}

\begin{figure}
	\includegraphics[width=0.5\textwidth]{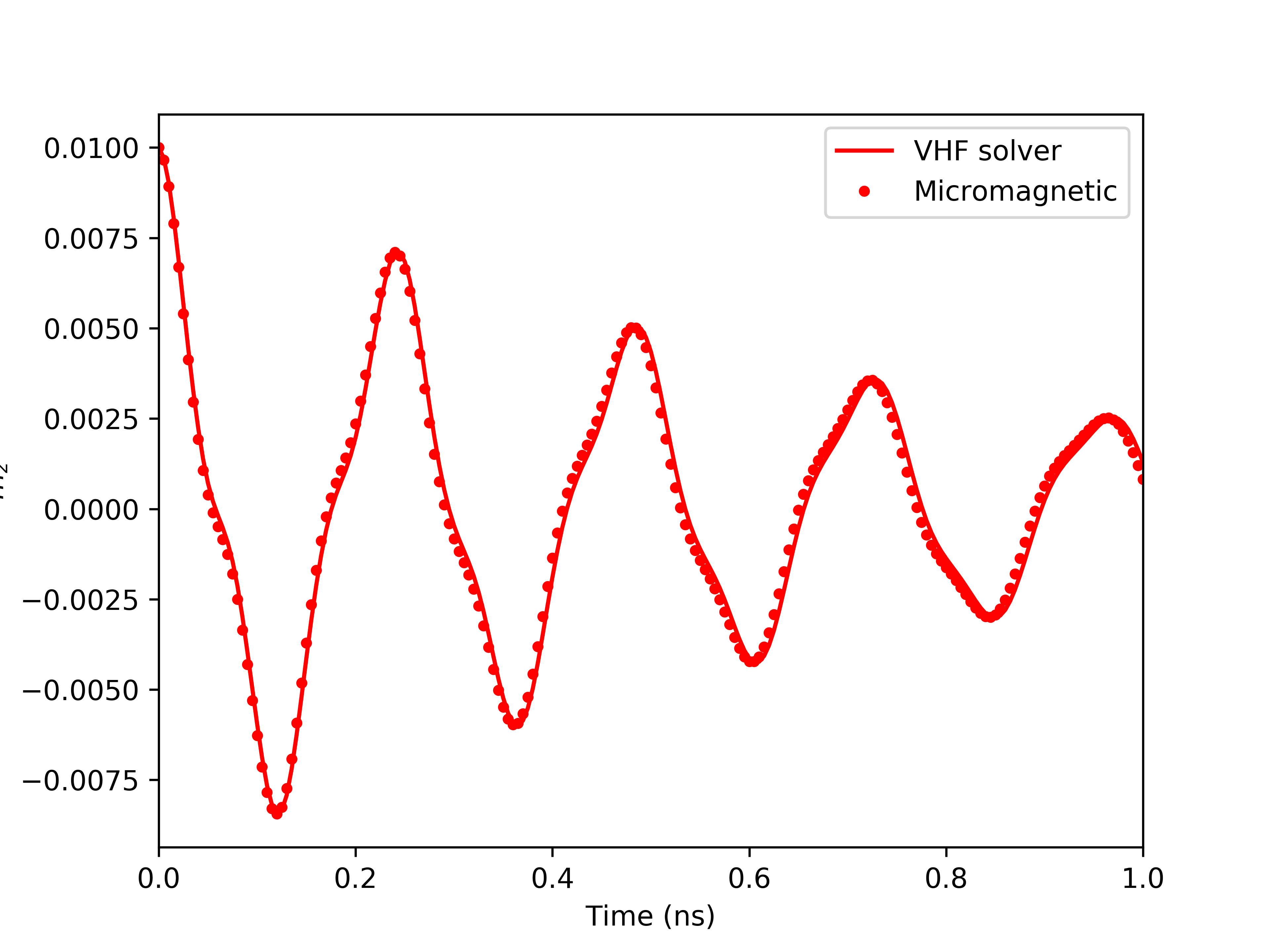}
	\label{f:decay-mz}
	\caption{Comparison of time dependence of averaged $m_z$ magnetization component calculated using direct micromagnetic simulations (points) and VHF-based calculations (solid line). Initial magnetization deviation was uniform in space and equal to $\vec{\delta m} = 0.01\vec{y}$.}
\end{figure}

Figures~\ref{f:decay-my} and \ref{f:decay-mz} show simulation results for $m_y(t)$ and $m_z(t)$ magnetization components, respectively, calculated using full-scale micromagnetic solver (points) and the developed linear VHF solver (solid lines). Initial magnetization distribution was uniform in space and equal to $\vec{\delta m} = 0.01\vec{y}$ (magnetizatoin was rotated towards the $y$ axis for about 2 degrees). As one can see from Fig.~\ref{f:decay-my} and \ref{f:decay-mz}, the VHF provides results that are practically indistinguishable from full-scale micromagnetic simulations. A minute shift of two sets of data at later times $t \approx 1$~ns is explained by the nonlinear frequency shift, which is present even at such small magnetization precession angles. This effect is not described by the \emph{linear} VHF approach (see the next subsection for comparison of nonlinear magnetization dynamics).

Note, that the time profiles of $m_y(t)$ and $m_z(t)$ noticeably deviate from a simple harmonic behavior, which is due to excitation of several modes in this numerical experiment. The developed VHF approach correctly describes the amplitude and phase relations between the excited modes.

\begin{figure}
	\includegraphics[width=0.5\textwidth]{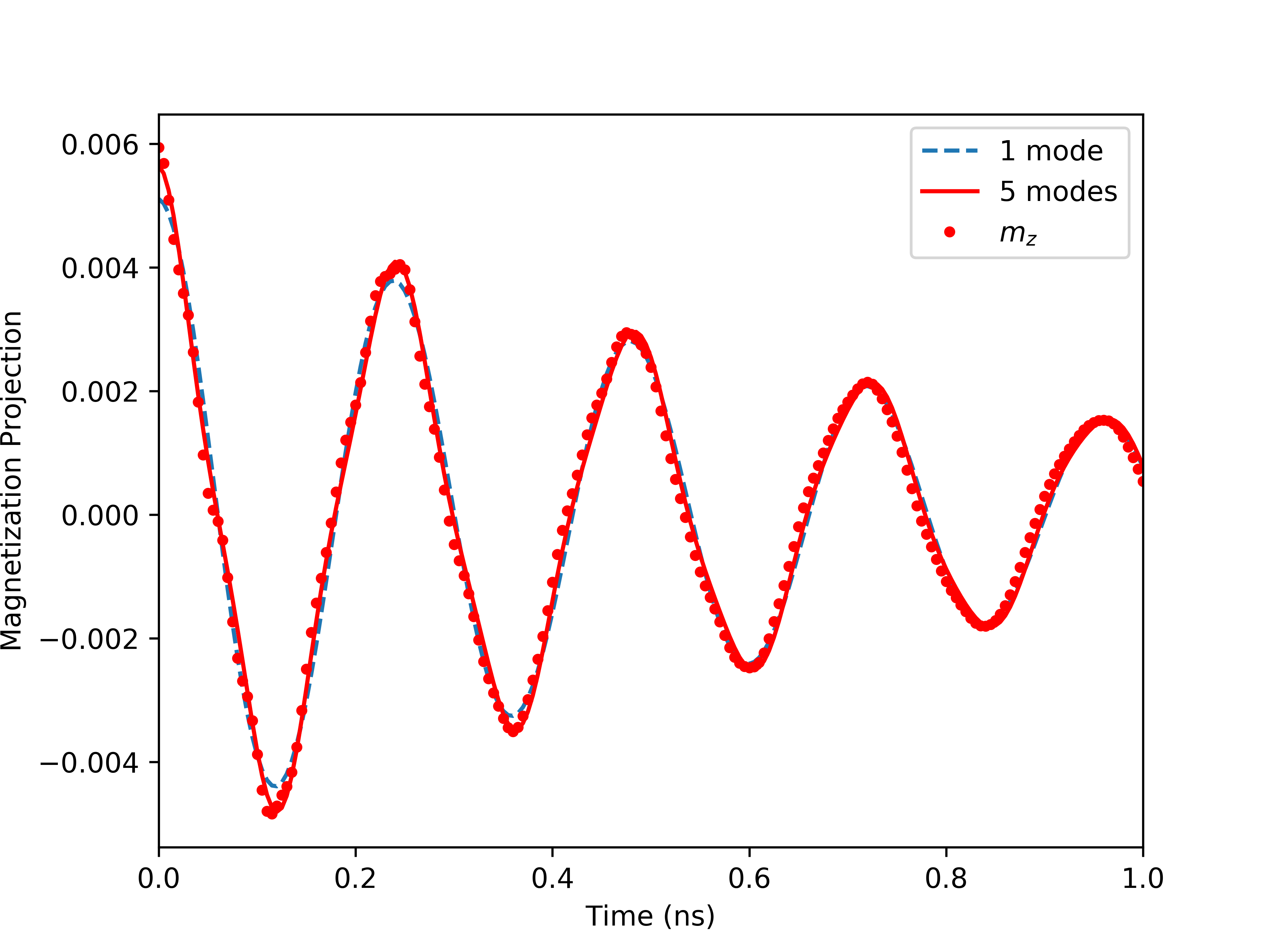}
	\label{f:decay-chess}
	\caption{Comparison of time dependence of averaged $m_z$ magnetization component calculated using direct micromagnetic simulations (points) and VHF-based calculations with $n=1$ (dashed blue line) and $n=5$ (solid red line) modes. The initial magnetization deviation from the ground state had the form of $3 \times 3$ chessboard profile with magnitude $|\vec{\delta m}| = 0.01$.}
\end{figure}

We have repeated the same numerical experiments for several different profiles of the initial magnetization deviation $\vec{\delta m}(\vec{r})$. As an example, we show in Fig.~\ref{f:decay-chess} results of numerical experiment with $3 \times 3$ chessboard initial deviation. One can see that the agreement between the full-scale micromagnetic simulations and VHF-based calculations is as good as in the previous example with uniform initial distribution. We would like to stress one more time, that this VHF-based experiment used the same information on spin wave profiles, frequencies, and damping rates, as the previous one, so the added calculation time was of the order of few milliseconds. We obtained the same excellent agreement between two simulation method for all studied initial distributions of the non-equilibrium magnetization, and for all cases the satisfactory results were obtained with not more than $n=5$ spin wave modes.

\subsection{Nonlinear Ferromagnetic Resonance}
\label{ss:nonlinear-fmr}

We have used the same test magnetic system to perform numerical experiments on nonlinear ferromagnetic resonance (FMR). We used the following procedure. First, we set the magnetic state of the test system to the ground state $\vec{m}_0(\vec{r})$. Then, we run simulations with microwave field (amplitude $h_{rf}$, frequency $f_{rf}$) switched on for 20~ns, which was enough to reach a steady state precession. After that, we run simulations for one additional period of the microwave magnetic field and found the magnitude of the spatially-averaged $\langle m_y\rangle$ component of dynamic magnetization. For convenience, we present below the simulation results as the values of the magnetic susceptibility $\chi_{yy} = M_s \langle m_y \rangle/h_{rf}$.

\begin{figure}
	\includegraphics[width=0.5\textwidth]{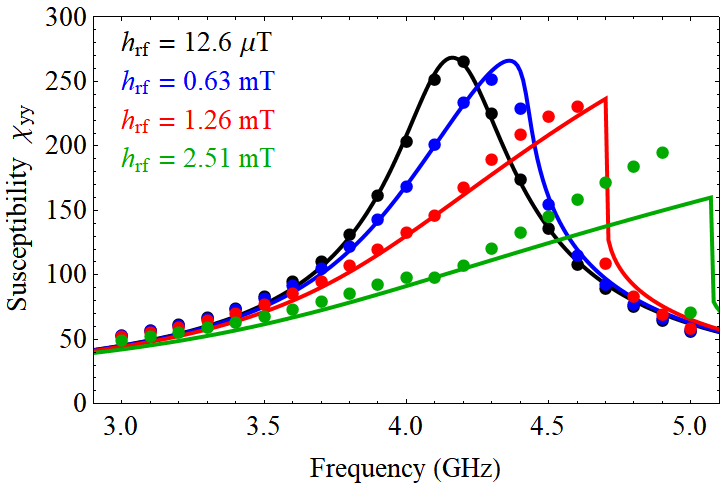}
	\label{f:nFMR}
	\caption{Comparison of the FMR response of the test magnetic system calculated using direct micromagnetic simulations (points) and VHF-based calculations (solid line) for different driving field magnitudes $h_{rf}$.}
\end{figure}

The described above numerical experiment was performed for several frequencies $f_{rf}$ in the range from 3~GHz to 5~GHz and several amplitudes of the microwave field from $\mu_0 h_{rf} = 12.6~\mathrm{\mu T}$ (linear regime) to $\mu_0 h_{rf} = 2.51~\mathrm{mT}$ (strongly nonlinear regime). The comparison of the results obtained using a standard full-scale micromagnetic solver and using the VHF-based analysis are shown in Fig.~\ref{f:nFMR}.

Black dots and line in Fig.~\ref{f:nFMR} correspond to the small magnitude of the driving microwave field $\mu_0 h_{rf} = 12.6~\mathrm{\mu T}$. For such small-amplitude excitations the magnetization dynamics is linear with very high accuracy. One can see from Fig.~\ref{f:nFMR} that both full-scale and VHF-based approaches correctly reproduce linear resonance curve with resonance frequency corresponding to the lowest spin wave mode at $f = 4.1$~GHz.

With the increase of the driving field magnitude to $\mu_0 h_{rf} = 0.63~\textrm{mT}$ (blue dots and line in Fig.~\ref{f:nFMR}), one can clearly see nonlinear distortions of the FMR curve, both in the shape of the curve and position of the maximum. Similarly to the previous case, VHF-based and full-scale approaches practically coincide. This proves the validity of the VHF approach to simulations of weakly-nonlinear magnetization dynamics. We would like to stress, that VHF results shown in Fig.~\ref{f:nFMR} were obtained without a single fitting parameter.

Red and green dots and lines in Fig.~\ref{f:nFMR} show the nonlinear FMR curves at even higher values of the driving field magnitude, $\mu_0 h_{rf} = 1.26~\mathrm{mT}$ and $\mu_0 h_{rf} = 2.51~\mathrm{mT}$, respectively. Such large amplitudes of the microwave magnetic field are hardly accessible experimentally, and the purpose of these simulations was to find a point at which the perturbative VHF approach starts to deviate from full-scale simulations. One can see, that the VHF-based simulations provide rather accurate quantitative description for the case $\mu_0 h_{rf} = 1.26~\textrm{mT}$ (red dots and line), but are only qualitatively correct for the case $\mu_0 h_{rf} = 2.51~\textrm{mT}$ (green). In the latter case, the average value of the $y$-component of dynamic magnetization was $\langle m_y \rangle = 0.5$, which corresponds to $30^\circ$ \emph{average} precession angle (the local precession angles were substantially larger). Thus, the perturbative VHF approach is quantitatively correct up to precession angles of about $30^\circ$, which is much larger than typical precession angles in majority of experiments.

\section{Conclusions}
\label{s:conclusions}

In conclusion, we developed a new approach to investigation of a weakly-nonlinear magnetization dynamics -- vector Hamiltonian formalism (VHF). The VHF is based on a vector transformation of a sphere to a plane (azimuthal Lambert projection), which preserves both the Hamiltonian structure and vector character of the Landau-Lifshits equation of magnetization dynamics. We derived simple and compact expressions for various nonlinear interaction coefficients of spin wave modes in the form of nonlinear ``overlap integrals'' of modes' profiles. The developed formalism is well-suited for hybrid analysis of magnetization dynamics, in which information about the linear dynamics of the studied magnetic system (eigen-frequencies and spin wave mode profiles) is obtained from numerical simulations, while nonlinear dynamics is analyzed based on quasi-Hamiltonian equations for spin wave amplitudes. The comparison of the results obtained using this method with results of full-scale nonlinear micromagnetic simulations demonstrates a very good agreement for the magnetization precession angles of up to at least $30^\circ$.

\appendix

\section{Mathematical Properties of the Linear Eigenproblem}
\label{app:Eigenproblem}

Here we will consider some basic mathematical properties of the linear eigenproblem Eq.~(\ref{eigenproblem}) in the case when the linear Hamiltonian of the system $\op{H}_0$ is a positive-definite operator. In this case, the operator $\op{H}_0$ can be represented as
\begin{equation}\label{cholesky}
	\op{H}_0 = \op{A}^+ \cdot \op{A}
\,,\end{equation}
where $\op{A}$ is a certain lower triangular matrix with real and positive diagonal entries and $\op{A}^+$ denotes Hermitian conjugate of $\op{A}$. Such decomposition of a Hermitian positive-definite operator is known as the Cholesky decomposition.

It should be noted, that finding the Cholesky decomposition Eq.~(\ref{cholesky}) is not technically much easier than solving the original eigenproblem Eq.~(\ref{eigenproblem}). Respectively, the aim of this section is not to provide any technical receipts for solving Eq.~(\ref{eigenproblem}), but to rigorously prove important mathematical properties of it. Another important note is that the operator $\op{H}_0$ is positive definite (in the case when the ground state $\vec{m}_0$ corresponds to a minimum of energy) only for vectors $\vec{s}$, which are orthogonal to $\vec{m}_0$. For ``longitudinal'' vectors $f(\vec{r}) \vec{m}_0(\vec{r})$ (which are parallel, in every point of space, to the local direction of $\vec{m}_0$), the action of $\op{H}_0$ is zero, $\op{H}_0 \cdot (f \vec{m}_0) = 0$, as it is obvious from the definition of this operator Eq.~(\ref{def-opH0}). Therefore, throughout this section we assume that the SEVs $\vec{s}$ are described using two-dimensional coordinate approach (as it is explained in Sec.~\ref{ss:SEV}), in which operator $\op{H}_0$ is positive definite, operator $\op{L}_0$ is invertible, and the projection operator $\op{P}_0$ is equivalent to the identity operator.

Using the Cholesky decomposition Eq.~(\ref{cholesky}), the eigenproblem Eq.~(\ref{eigenproblem}) takes the form
\begin{equation}\nonumber
	-i \omega_\alpha \op{L}_0 \cdot \vec{s}_\alpha = \op{A}^+ \cdot \op{A} \cdot \vec{s}_\alpha
\,.\end{equation}
Multiplying both sides of this equation by $(\op{A}^{-1})^+$ (note, that the operator $\op{A}$ is always invertible) and introducing new eigenvectors
\begin{equation}\label{aux-vecu}
	\vec{u}_\alpha = \op{A} \cdot \vec{s}_\alpha
\end{equation}
this equation can be reformulated as
\begin{equation}\label{aux-eigenproblem-cholesky}
	\lambda_\alpha \vec{u}_\alpha = \op{B} \cdot \vec{u}_\alpha
\,,\end{equation}
where $\lambda_\alpha = 1/\omega_\alpha$ and
\begin{equation}\label{def-opB}
	\op{B} = -i (\op{A}^{-1})^+ \cdot \op{L}_0 \cdot \op{A}^{-1}
\,.\end{equation}

As one can easily see, the operator $\op{B}$ is a Hermitian operator, so the reformulated eigenproblem Eq.~(\ref{aux-eigenproblem-cholesky}) is a standard Hermitian eigen-value problem. Respectively, the set of eigenvectors $\vec{u}_\alpha$ (and, respectively, set of vectors $\vec{s}_\alpha$) forms a complete set of vector functions, orthogonal to the ground state $\vec{m}_0$, and all eigenvalues $\lambda_\alpha$ (and eigenfrequencies $\omega_\alpha = 1/\lambda_\alpha$) are real.

Moreover, the eigenvectors $\vec{u}_\alpha$ that correspond to different eigenvalues $\lambda_\alpha$ are orthogonal to each other, and multiple eigenvectors corresponding to a degenerate eigenvalue can be mutually orthogonalized. Thus,
\begin{equation}\label{aux-orthogonality}
	\int_{V_s} \vec{u}_\alpha^+ \cdot \vec{u}_{\alpha'} \, d\vec{r} = \epsilon_\alpha \Delta_{\alpha, \alpha'}
\,,\end{equation}
where $\epsilon_\alpha$ are certain positive normalization constants and $\Delta_{\alpha, \beta}$ is the Kronecker delta.

Using the definition of the auxiliary vectors $\vec{u}_\alpha$ Eq.~(\ref{aux-vecu}), the orthogonality condition Eq.~(\ref{aux-orthogonality}) can be reformulated in terms of the SEVs $\vec{s}_\alpha$:
\begin{equation}\label{aux-orthogonality-s}
	\int_{V_s} \vec{s}_\alpha^+ \cdot \op{H}_0 \cdot \vec{s}_{\alpha'} \, d\vec{r} = \epsilon_\alpha \Delta_{\alpha, \alpha'}
\,.\end{equation}
This condition is equivalent to Eq.~(\ref{ortho-energy}) with $\epsilon_\alpha = \hbar_\alpha \omega_\alpha$, and can also be derived in a slightly less rigorous way directly from the eigenproblem Eq.~(\ref{eigenproblem}). The presented above derivation also proves that the product of the mode's norm $\hbar_\alpha$ and its eigenfrequency $\omega_\alpha$ is always a positive quantity (for a ground state $\vec{m}_0$ that corresponds to a minimum of energy), \ie that the modes with positive norms have positive eigenfrequencies.

Another standard property of the Hermitian eigenproblem Eq.~(\ref{aux-eigenproblem-cholesky}) is that the operator $\op{B}$ is diagonal in the basis of eigenvectors $\op{u}_\alpha$, which can be written as another orthogonality condition:
\begin{equation}
	\int_{V_s} \vec{u}_\alpha^+ \cdot \op{B} \cdot \vec{u}_{\alpha'} \, d\vec{r} 
		= \epsilon_\alpha \lambda_\alpha \Delta_{\alpha, \alpha'}
\,.\end{equation}
This condition, also, can be rewritten in terms of the original SEVs $\vec{s}_\alpha$:
\begin{equation}
	\int_{V_s} \vec{s}_\alpha^+ \cdot \op{L}_0 \cdot \vec{s}_{\alpha'} \, d\vec{r} 
		= i \epsilon_\alpha \lambda_\alpha \Delta_{\alpha, \alpha'}
\,,\end{equation}
and is equivalent to Eq.~(\ref{orthogonality}).

\section{Elimination of Non-Resonant Three-Magnon Processes}
\label{app:Elimination}

Here we shall briefly describe the procedure of elimination of three-magnon processes $\H_3$ in the case when such processes are non-resonant, \ie the condition Eq.~(\ref{nonresonant-3}) is satisfied. Consider a weakly-nonlinear transformation of the spin wave amplitudes $c_\alpha$:
\begin{equation}\label{elim:transform-1}
	c_\alpha \to c_\alpha' = c_\alpha + \sum_{\beta\gamma} A_{\alpha,\beta\gamma} c_\beta c_\gamma + \ldots
\,.\end{equation}
If this transformation is canonical, \ie preserves the form of the Poisson brackets Eq.~(\ref{def-Poisson}), the equations of motion for the new amplitudes $c_\alpha'$ will have the same form Eq.~(\ref{eqmot-Poisson}), where the Hamiltonian function $\H$ should be written using the transformed amplitudes and will have a different functional form. Using properly chosen transformation coefficients, one may simplify the transformed Hamiltonian, in particular, eliminate the three-magnon term $\H_3$ if it is non-resonant. 

The eliminated non-resonant processes lead, in second perturbation order, to renormalization of coefficients of higher-order ($\H_4$) resonant processes, \ie to renormalization of coefficients $W_{\alpha\beta\gamma\delta} \to W_{\alpha\beta\gamma\delta}' = W_{\alpha\beta\gamma\delta} + \Delta W_{\alpha\beta\gamma\delta}$. In the case of magnetic systems, the correction $\Delta W_{\alpha\beta\gamma\delta}$ can be, in general, of the same order of magnitude as the original interaction coefficient $W_{\alpha\beta\gamma\delta}$ and, strictly speaking, cannot be ignored. To calculate the correction $\Delta W_{\alpha\beta\gamma\delta}$, one has to use weakly-nonlinear canonical transformation Eq.~(\ref{elim:transform-1}) explicitly taking into account both quadratic and cubic terms in the expansion $c_\alpha'(c_\beta)$, which leads to rather cumbersome and technically difficult expressions.

Therefore, instead of using explicit form of the transformation Eq.~(\ref{elim:transform-1}), we will employ the fact that any Hamiltonian dynamics described by equations of the form Eq.~(\ref{eqmot-Poisson}) is itself a canonical transformation. Then, we can consider a canonical transformation generated by certain ``Hamiltonian function'' $\F$:
\begin{equation}\label{elim:transform-2}
	c_\alpha' = c_\alpha + [\F, c_\alpha] + \frac12 \, [\F, [\F, c_\alpha]] + \ldots
\,.\end{equation}
Choosing $\F$ as a cubic function in spin wave amplitudes,
\begin{equation}
	\F = \frac16 \sum_{\alpha\beta\gamma} F_{\alpha\beta\gamma} c_\alpha c_\beta c_\gamma
\end{equation}
leads to the desired weakly-nonlinear behavior Eq.~(\ref{elim:transform}), while the ``Poisson bracket'' form of the transformation Eq.~(\ref{elim:transform-2}) guarantees that it is a canonical one.

The transformed Hamiltonian $\H'(c_\alpha') = \H(c_\alpha)$ can also be written as a ``Poisson-bracket expansion'':
\begin{equation}
	\H' = \H - [\F, \H] + \frac12 [\F, [\F, \H]] + \ldots
\,.\end{equation}

Using weakly-nonlinear expansions of $\H = \H_2 + \H_3 + \H_4$ and $\H' = \H_2' + \H_3' + \H_4'$, one can relate different-order terms in the original $\H$ and transformed $\H'$ Hamiltonian functions:
\begin{subequations}
\begin{eqnarray}
	\H_2' &=& \H_2
\,,\\
	\H_3' &=& \H_3 - [\F, \H_2]
\,,\\
	\H_4' &=& \H_4 - [\F, \H_3] + \frac12 \, [\F, [\F, \H_2]]
\,.\end{eqnarray}
\end{subequations}
Thus, weakly-nonlinear transformation leaves the quadratic part of the Hamiltonian $\H_2$ unchanged. The three-magnon term $\H_3'$ vanishes if $\F$ satisfies
\begin{equation}\label{elim:cond}
	[\F, \H_2] = \H_3
\,,\end{equation}
in which case $\H_4'$ can be written in a very simple form
\begin{equation}\label{elim:H4prim}
	\H_4' = \H_4 - \frac12 , [\F, \H_3]
\,.\end{equation}

Direct evaluation of $[\F, \H_2]$ gives
\begin{equation}
	[\F, \H_2] = \frac{i}{6} \sum_{\alpha\beta\gamma} (\omega_\alpha + \omega_\beta + \omega_\gamma)
		F_{\alpha\beta\gamma} c_\alpha c_\beta c_\gamma
\,.\end{equation}
The elimination condition Eq.~(\ref{elim:cond}) requires
\begin{equation}\label{elim:F}
	F_{\alpha\beta\gamma} = -i \, 
		\frac{V_{\alpha\beta\gamma}}{\omega_\alpha + \omega_\beta + \omega_\gamma}
\,,\end{equation}
which can be satisfied if all three-magnon processes are non-resonant Eq.~(\ref{nonresonant-3}).

The transformed four-magnon Hamiltonian $\H_4'$ Eq.~(\ref{elim:H4prim}) has the form
\begin{equation}
	\H_4' = \H_4 -\frac{i}{8} \, \sum_{\alpha\beta\gamma\delta\epsilon}
		\frac{F_{\alpha\beta\epsilon}V_{\epsilon^*\gamma\delta}}{\hbar_\epsilon}
		c_\alpha c_\beta c_\gamma c_\delta
\,.\end{equation}
Using the expression Eq.~(\ref{elim:F}) for coefficients $F_{\alpha\beta\gamma}$, one can rewrite $\H_4'$ in the form
\begin{equation}
	\H_4' = \frac{1}{24} \, \sum_{\alpha\beta\gamma\delta} W_{\alpha\beta\gamma\delta}' 
		c_\alpha c_\beta c_\gamma c_\delta
\end{equation}
with renormalized coefficients $W_{\alpha\beta\gamma\delta}'$ given by the symmetrized expressions Eq.~(\ref{W-renormalized}).


\end{document}